\documentclass[12pt,a4paper,tightenlines,nofootinbib]{revtex4}

\usepackage{amssymb}
\usepackage{amsmath}
\usepackage{epsf}
\usepackage{psfrag}

\begin{document}

\author{Ariel M\'egevand}
\email[Electronic address: ]{megevand@ifae.es} \affiliation{Departament de F{\'\i}sica
Fonamental, Universitat de Barcelona, Diagonal 647, 08028 Barcelona, Spain and \\
IFAE, Universitat Aut{\`o}noma de Barcelona, 08193 Bellaterra (Barcelona), Spain}
\title{First-order cosmological phase transitions in the radiation dominated
era}
\date{\today}

\begin{abstract}
We consider first-order phase transitions of the Universe in the radiation-dominated
era. We argue that in general the velocity of interfaces is non-relativistic due to
the interaction with the plasma and the release of latent heat. We study the general
evolution of such slow phase transitions, which comprise essentially a short reheating
stage and a longer phase equilibrium stage. We perform a completely analytical
description of both stages. Some rough approximations are needed for the first stage,
due to the non-trivial relations between the quantities that determine the variation
of temperature with time. The second stage, instead, is considerably simplified by the
fact that it develops at a constant temperature, close to the critical one. Indeed, in
this case the equations can be solved exactly, including back-reaction on the
expansion of the Universe. This treatment also applies to phase transitions mediated
by impurities. We also investigate the relations between the different parameters that
govern the characteristics of the phase transition and its cosmological consequences,
and discuss the dependence of these parameters with the particle content of the
theory.
\end{abstract}

\maketitle

\section{Introduction}

It is well known that the Universe could have undergone several phase transitions in
the early stages of its history, most of them associated with the spontaneous symmetry
breaking of some symmetry. Some examples are the quark-hadron phase transition at the
QCD scale, the phase transitions associated to the electroweak $SU(2)\times U(1)$
symmetry breaking or to Grand Unified Theories, and the Peccei-Quinn phase transition,
related to the axion field and the strong CP problem. Cosmological phase transitions
generically produce cosmic relics, such as topological defects, magnetic fields, or
baryon number asymmetries, with potentially important cosmological consequences. The
mechanisms for generating these relics build on the dynamics of the phase transition.

In a first-order phase transition the dynamics is essentially determined by the
nucleation and expansion of bubbles. At zero temperature, when a true vacuum bubble
nucleates, it rapidly begins to expand with almost the velocity of light \cite{c77}.
On the contrary, at high temperature, the bubble expands in a hot plasma, which is
perturbed by the motion of the bubble walls. The plasma thus opposes a resistance to
the expansion, that depends on the wall velocity. As a consequence, bubble walls feel
a friction force, which prevents them to accelerate indefinitely. Then, the velocity
quickly reaches a stationary value, determined by the viscosity of the plasma and the
pressure difference between the low temperature phase and the supercooled one. These
quantities depend on the model, and it is known that the friction can be large enough
to prevent the wall from acquiring relativistic velocities \cite{mp95,js01}.
Furthermore, the release of latent heat at the interfaces of the phase transition
reheats the surrounding plasma up to a temperature that in most cases is close to the
critical temperature. Consequently, the pressure difference that drives bubble
expansion may decrease considerably, causing a drastic slowdown of the phase
transition \cite{hkllm93,h95}. Therefore, at the radiation dominated era bubble walls
generically undergo non-relativistic motion. Since the heat liberated at the
interfaces is taken away by sound waves, the temperature can be assumed to be
homogeneous, which simplifies the analysis.

In order to have a first-order phase transition, the free energy must allow the
coexistence of two phases. Therefore, we will assume that the free energy density
$\mathcal{F}$ depends on some order parameter (usually a Higgs field) $\phi \left(
x\right) $. In a certain range of temperatures, the free energy bears two minima
separated by a barrier; one of them at $\phi =0$, which corresponds to the symmetric
phase, and the other at a nonzero value $\phi _{m}\left( T\right) $, corresponding to
the broken symmetry phase. The difference in free energy density $V\left( \phi
,T\right) $ between some value $\phi $ of the order parameter and $\phi =0$, is
generally given by a finite-temperature effective potential. The free energy
difference between the two minima is thus given by
\begin{equation}
\mathcal{F}_{b}-\mathcal{F}_{u}= V\left( \phi _{m}\left( T\right) ,T\right) \equiv
V\left( T\right). \label{deltav}
\end{equation}
At the critical temperature the two minima are degenerate, $V\left( \phi _{m}\left(
T_{c}\right) ,T_{c}\right) =0$. Above the critical temperature the symmetric phase is
the stable one, while below $T_{c}$ it becomes metastable, being $\phi_m$ the absolute
minimum. Finally, at some temperature $T_{0}<T_{c}$, the barrier between the minima
disappears, and the symmetric phase becomes unstable. At some stage between the
temperatures $T_{c}$ and $T_{0}$, bubbles of broken-symmetry phase will be formed in
the sea of symmetric phase. A bubble can be described as a configuration in which the
order parameter is non-vanishing inside a spherical region (see e.g. \cite{q99}).
After being nucleated, a bubble will grow with a velocity that depends on the pressure
difference at the interface, $\Delta p=-V\left( T\right) $, and on the viscosity of
the hot plasma in which it expands.

In this work we will be concerned with first-order phase transitions in the
radiation-dominated epoch.  Our aim is to study the development of such phase
transitions within a completely analytical approach. Here we concentrate on the
determination of the parameters that govern relic formation (e.g. number density of
bubbles, bubble wall velocity, etc.). An analytic study of cosmological consequences
will be addressed in \cite{cosmo}, where the results of the present analysis will be
used. In Section \ref{remnants} we briefly review the influence of phase transition
dynamics on the mechanisms for generating cosmic remnants in first-order phase
transitions.

In section \ref{phaseeq} we study a phase transition that completely develops at
$T=T_{c}$, with the two phases in equilibrium. This is a good approximation in the
case of inhomogeneous nucleation in the presence of impurities. In the case of
homogeneous nucleation of bubbles, this approximation correctly describes the
evolution of the phase transition after some latent heat has been released. An
interesting feature of phase equilibrium is that it is simple enough to solve
analytically, including back-reaction on the expansion of the Universe. This is due to
the fact that temperature is constant all the way through the phase transition. We
thus can obtain the fraction of volume of the Universe that is occupied by the
low-temperature phase as a function of time, with no need of any numerical
calculations.

Section \ref{superc} is devoted to the analysis of the phase
transition in the case of homogeneous nucleation. The main
difference with the previous case is the initial stage of
supercooling and quick reheating back to the critical temperature.
Section \ref{transicion} contains an analytical study of the phase
transition dynamics. In sections \ref{physquant} and \ref{perturb}
we analyze the relations between the different physical parameters
involved in the dynamics of the phase transition, leaving some
technical discussions to the appendices. Our conclusions are
summarized in section \ref{conc}.

\section{Cosmological consequences of phase transitions \label{remnants}}

Several cosmological objects may be formed in a phase transition of the Universe.
Their abundance and characteristics depend on details of the development of the phase
transition. Due to the complexity of the mechanisms by which these objects are created
and the difficulty of describing the phase transition, several details of the dynamics
(e.g. the variation of the nucleation rate or the wall velocity during the transition)
are often disregarded for the sake of simplicity. An analytical study of the phase
transition is thus important since analytical expressions will help taking into
account the dynamics in a more rigorous way. In this section we review how phase
transition dynamics affects the cosmological remnants.

\subparagraph{Electroweak baryogenesis.}

During the last two decades there has been much interest in the
possibility that the electroweak phase transition could be the
framework for the generation of the baryon number asymmetry of the
Universe (BAU). A first-order electroweak phase transition
provides the three Sakharov's conditions for the generation of a
BAU, although physics beyond the minimal Standard Model (SM) is
mandatory in order to obtain a quantitatively satisfactory result
(for reviews on electroweak baryogenesis see \cite{ckn93}). Due to
$CP$ violating interactions of particles with the bubble walls, an
asymmetry between left handed quarks and their antiparticles is
generated in front of the walls of expanding bubbles. This
asymmetry biases the baryon number violating sphaleron processes
in the symmetric phase. The resulting baryon asymmetry is caught
by the walls and enter the bubbles, where baryon number violating
processes are turned off.

It is important that the sphaleron processes be suppressed in the broken
symmetry phase, in order to avoid the washout of the generated BAU when
equilibrium is established after the phase transition. This requirement
imposes a condition on the value of the Higgs field at the temperature of
the transition \cite{s87}, $\phi _{m}\left( T_{t}\right) /T_{t}\gtrsim 1$.
Since $\phi _{m}$ is the order parameter, this is a condition on the
strength of the first-order phase transition.

The resulting BAU depends also on the bubble wall velocity. On one hand, if the
velocity is too large, the left-handed density perturbation will pass so quickly
through a given point in space that sphaleron processes will not have enough time to
produce baryons; thus the resulting BAU will be small. On the other hand, for very
small velocities thermal equilibrium will be restored and the baryon asymmetry will be
erased by sphalerons; thus the BAU will be small again. As a consequence, the
generated baryon number will have a peak at a given wall velocity, which is of order
$10^{-2}$ \cite{lmt92,ckn92,ck00}.

Both values of $\phi _{m}$ and $v_{w}$ depend on $T$ and vary during the phase
transition. The dependence of the wall velocity is more critical, since reheating may
cause it to descend two orders of magnitude before the transition completes.
Baryogenesis may be either enhanced or suppressed by this effect \cite{h95,m01},
depending on which side of the peak the initial velocity lies.

\subparagraph{Baryon inhomogeneities.}

A general feature of cosmological phase transitions is the difference of particle
masses between the high- and low-temperature phases. These mass differences give rise
to different number densities in the two phases. At the QCD phase transition, for
instance, baryons are much heavier in the hadron phase than in the deconfined quark
phase. As the hadron phase expands, baryons are pushed into the quark phase region,
leading to localized clumps of high density surrounded by large voids of low baryon
density \cite{w84,ah85,fma88}. An important consequence is that large amplitude, small
scale density fluctuations may survive until the nucleosynthesis epoch, affecting the
standard scenario of big bang nucleosynthesis. Therefore, inhomogeneous
nucleosynthesis may put constraints on the quark-hadron phase transition (see e.g.
\cite{ahs87}). Moreover, if the quark phase reaches sufficiently high density, its
pressure may balance that of the hadron phase. The quark matter trapped in small
regions of space forms quark plasma objects that may survive until the present epoch
\cite{w84,fj84}.

Baryon inhomogeneities may also arise at the electroweak phase
transition, since the amount of baryons produced through
electroweak baryogenesis depends drastically on the wall velocity,
and the latter has a considerable variation during the phase
transition \cite{h95}. The geometry of the electroweak
inhomogeneities is in general quite different from the QCD case.
If the BAU peaks at a certain wall velocity, then the high density
regions will form spherical walls, whose radius depends on the
moment in the bubble evolution at which the peak velocity is
attained. Furthermore, baryon number densities with the wrong sign
may arise in some regions of space, depending on the baryogenesis
scenario \cite{gs98,ck00,m01}. This gives rise to the interesting
possibility of nucleosynthesis in the presence of antibaryons (see
for example \cite{rj01}).

However, due to baryon diffusion and ``neutrino inflation", baryon
inhomogeneities generated at the electroweak phase transition
hardly survive until the nucleosynthesis time (see e.g.
\cite{h95,bdr95}). Nevertheless, they may survive until the QCD
scale. In that case, electroweak scale inhomogeneities can act as
impurities for the quark-hadron phase transition (see next
section).

In summary, we can say that the amplitude and scale of baryon inhomogeneities
generically depend on the mean nucleation distance and on the variation of the
velocity of bubble expansion.

\subparagraph{Topological defects and magnetic fields.}

If a global $U\left( 1\right) $ symmetry is spontaneously broken
at a first-order phase transition, the phase angle $\theta $ of
the Higgs field within each nucleated bubble is essentially
constant, but phases in different bubbles are uncorrelated. When
bubbles collide, the discontinuity in the Higgs phase is smoothed
out to become a continuous variation. The so called "geodesic
rule" states that (for energetic reasons) the shortest path
between the two phases is chosen \cite{k76}. When three bubbles
meet, a vortex (in two spatial dimensions) or a string (in 3d) may
be trapped between them. This mechanism is obviously generalized
to higher symmetry groups and other kinds of topological defects.
Ignoring the dynamics of phase equilibration, it is easy to see
that the number density of defects is proportional to the number
density of bubbles. However, the final number of defects will
depend strongly on the velocity of bubble expansion \cite{kv95}.
If the latter is much less than the velocity of light, then phase
equilibration between two bubbles will have probably completed
before they encounter a third bubble, thus reducing the chances of
trapping a string.

The above picture is in fact a rough simplification of the defect-formation
problem. One complication is due to dissipation, since the Higgs field is
coupled to the other fields in the thermal bath. Another complication arises
when considering a gauge symmetry, and is caused by the fact that the phase
of the Higgs field is not a gauge-invariant quantity, so it is convenient to
define a gauge-invariant phase difference between two bubbles \cite{kv95}.
The phase difference is thus linked to the gauge field. In this case,
dissipation can be taken into account by introducing the conductivity of the
plasma. Then one can model the collision of three bubbles and calculate the
evolution of the phase difference and gauge field. One can say that a vortex
is formed whenever a quantum of magnetic flux is trapped in the
unbroken-symmetry region between the three bubbles.

From the above it is clear that the formation of local vortices is
associated with the generation of magnetic fields. Therefore
bubble collision constitutes also a mechanism for generating the
cosmic magnetic fields (e.g., see \cite{gr98}). Of course, the
magnetic field that is formed in this way corresponds to a
spontaneously-broken symmetry which cannot be the electromagnetic
$U\left( 1\right) _\textrm{em}$. Nevertheless, this mechanism can
take place at the electroweak phase transition, where unstable
cosmic strings \footnote{see for example \cite{ewstrings}.} and
hypermagnetic fields may be formed. The latter are subsequently
converted to $U\left( 1\right) _\textrm{em}$ magnetic fields. It
is interesting to note that the presence of magnetic fields may
affect the dynamics of the electroweak phase transition (see e.g.
\cite{eek98}).

Calculating the magnitude of the magnetic fields and the density of defects that are
left at the end of the phase transition involves the passage from three-bubble
collision simulations to the computation of the full phase transition. Evidently, this
is a difficult task. Although some simulations have been made (e.g., \cite{bkvv95}),
several simplifications are generally required, which include forgetting about
variations in the nucleation rate and the velocity of expansion of bubbles during the
transition. An analytical investigation of phase transition dynamics may therefore
clarify the picture and provide useful tools for the calculation of defect formation
and magnetic field generation.

\section{Phase equilibrium \label{phaseeq}}

We begin by considering the limiting case in which the first-order
phase transition is as slowest as possible, namely, that of
coexistence of the high- and low-temperature phases at the
critical temperature $T_{c}$. Such a cosmic separation of phases
has been studied for the QCD phase transition
\cite{s82,w84,ah85,eikr92}. At the critical temperature there is
no pressure difference between phases at the bubble walls, so the
bubble expansion takes place almost in equilibrium. Assume at
$T=T_c$ there are already regions with low-temperature phase. As
the Universe expands, the fraction of space occupied by these
regions increases, as the high-temperature phase converts to
low-temperature phase. The loss of energy due to the expansion of
the Universe is thus compensated by the latent heat released at
the interfaces, and the temperature remains constant. In this
scenario there is no supercooling.

Since at $T_{c}$ the nucleation rate vanishes, such a first-order
phase transition is only possible in the presence of impurities
that induce the formation of bubbles. In this case inhomogeneous
nucleation theory applies. In a phase transition mediated by
impurities there will still be some supercooling, which we neglect
in this section for simplicity. The role of impurities in the
early Universe could be played for instance by topological
\cite{s81,gw81} or non-topological solitons
\cite{fmm96,fgmm94,nontopol}. These may exist in the
high-temperature phase, containing the low-temperature phase in
their core. In this case their configurations become unstable or
metastable below the critical temperature. When the system cools
below $T_{c}$ these objects begin to expand and convert the
Universe into the true vacuum.

Another example of inhomogeneous nucleation is the case of the QCD
phase transition in the presence of baryon number inhomogeneities
\cite{cm96}. These may arise as a natural consequence of
electroweak baryogenesis \cite{h95}, and can survive until the QCD
scale \cite{jf94}. Since the critical temperature is different in
regions with different chemical potential \cite{hjssv98,Fodor},
bubbles will nucleate first in those regions with a higher $T_{c}$
\cite{s03}. If such regions are relatively small, and if they
achieve the necessary amount of supercooling while the surrounding
background reaches the critical temperature, then the baryon
inhomogeneities operate as impurities for inhomogeneous
nucleation.

Even if the phase transition proceeds by homogeneous nucleation of
bubbles, phase equilibrium will describe quite well a good part of
the transition, whenever the latent heat is at least comparable to
the energy density difference between the critical temperature and
that at which nucleation effectively begins. In that case, the
energy released will reheat the plasma back to a temperature very
close to $T_{c}$. At that moment bubble nucleation virtually stops
and the two phases remain close to equilibrium until the full
latent heat of the transition is eliminated. We will analyze this
case in the next section.

The customary equation for the adiabatic expansion,
$\dot{\rho}=-3H\left( \rho +p\right) $, tells us how the Universe
takes energy from the hot plasma. Here $\rho $ is the energy
density, $p$ is the pressure, and $H$ is the expansion rate. It
can be equivalently written in terms of entropy density,
$\dot{s}=-3Hs$, which is just the statement of entropy
conservation, $S=constant$. It will be convenient to use it in the
form $s\propto a^{-3}$. At the beginning of the phase transition
the whole Universe is in the symmetric phase, while at the end of
the transition it is filled with the broken symmetry one, so we
can write
\begin{equation} s=s_{u}\left(
T_{c}\right) a_{i}^{3}/a^{3}=s_{b}\left( T_{c}\right)
a_{f}^{3}/a^{3},  \label{entropy1}
\end{equation}
where $s_{u(b)}$ is the entropy density of the unbroken (broken)
symmetry phase, $a$ is the cosmic scale factor, and
$a_{i(f)}\equiv a\left( t_{i(f)}\right) $ its value at the
beginning (end) of the transition. We assume that, since the phase
transition occurs very slowly, the latent heat released at the
interfaces is quickly distributed throughout space and the
temperature is homogeneous. We also assume that pressure and
temperature remain constant during the phase transition. Phase
coexistence at $T=T_{c}$ means that there are regions of space
with different equations of state. Thus the entropy density has
different constant values $s_{b}\left( T_{c}\right) $ and
$s_{u}\left( T_{c}\right) $ inside and outside the bubbles of
broken symmetry phase respectively. The quantity $s$ in
Eq.~(\ref{entropy1}) is the average entropy density of the whole
system. The entropy in a comoving volume $V_{U}=V_{b}+V_{u}$ is
the sum of two contributions, $S=s_{b}V_{b}+s_{u}V_{u}$. Thus,
\begin{equation}
s=s_{u}+\left( s_{b}-s_{u}\right) f_{b},  \label{entropy2}
\end{equation}
where $f_{b}$ is the fraction of space that is already in the
broken-symmetry phase. The entropy, energy and pressure are
derived from the free energy. At $T=T_{c}$ the pressure $p_{c}$ is
the same in the two phases, $\Delta p=-V\left( T_{c}\right) =0$.
The latent heat of the phase transition is
\begin{equation}
L=\rho _{u}-\rho _{b}=T_{c}\left( s_{u}-s_{b}\right) =T_{c}V^{\prime }\left(
T_{c}\right) .  \label{latent}
\end{equation}
From Eqs.~(\ref{entropy1}) and (\ref{entropy2}), and using
(\ref{latent}), we obtain the dependence of $f_{b}$ on the scale
factor
\begin{equation}
f_{b}=\frac{T_{c}s_{u}}{L}\left( 1-\frac{a_{i}^{3}}{a^{3}}\right) .
\label{fractiontc}
\end{equation}

The dependence of the scale factor on time is given by the
Friedman equation
\begin{equation}
H^{2}\equiv \left( \frac{\dot{a}}{a}\right) ^{2}=\frac{8\pi G}{3}\rho ,
\label{friedman}
\end{equation}
where for simplicity we have neglected the term $k/a^{2}$. At
constant temperature $T_c$ and pressure $p_c$, the energy density
is given by
\begin{equation}
\rho =T_{c}s-p_{c}=T_{c}s_{u}a_{i}^{3}/a^{3}-p_{c},
\end{equation}
where we have used Eq.~(\ref{entropy1}) in the last equality.
Inserting $\rho$ in Eq.~(\ref{friedman}) and writing
$H=\frac{1}{3}a^{-3}da^{3}/dt$ we can easily solve the equation
for $a^{3}\left( t\right) $,
\begin{equation}
\left( \frac{a}{a_{i}}\right) ^{3}=\frac{T_{c}s_{u}}{p_{c}}\sin ^{2}\left[
\omega \left( t-t_{i}\right) +\delta \right] ,  \label{scale}
\end{equation}
where $\omega =\sqrt{6\pi Gp_{c}}$ and the constant phase is
determined by the initial condition $a\left( t_{i}\right) =a_{i}$,
$\delta =\arcsin \sqrt{p_{c}/T_{c}s_{u}}$.

During the phase transition we have two coexisting phases in the radiation dominated
era, and we may consider different possibilities for the equations of state. The
simplest one is to assume that the Universe is radiation-dominated before the phase
transition, i.e.,
\begin{equation}
p_{u}=\rho _{u}/3, \label{purad}
\end{equation}
with
\begin{equation}
\rho _{u}\left( T\right) =\pi ^{2}g_{\ast }T^{4}/30,  \label{rourad}
\end{equation}
where $g_{\ast }$ is the number of effectively massless species \footnote{In general
$g_{\ast }$ depends on $T$. We will assume for simplicity that $ g_{\ast }$ is
constant during the phase transition. We discuss the effect of a variation of $g_{\ast
}$ in Section \ref{physquant}.}. In fact, this is not a realistic situation. In the
symmetric phase the Higgs vev does not correspond to the true vacuum, so we should add
a constant energy density to account for the energy of this state and have a
negligible cosmological constant after the phase transition. It is interesting,
however, to consider first this simpler case. Hence, at $T=T_{c}$ Eqs. (\ref{purad})
and (\ref{rourad}) give $p_{c}=\rho _{u}/3$ and $T_{c}s_{u}=4\rho _{u}/3$, so
$T_{c}s_{u}/p_{c}=4$, $\delta =\pi /6$, and $\omega =\sqrt{3} H_{i} /2$.

Before the phase transition $H=\sqrt{8\pi G\rho _{u}/3}$, with
$\rho _{u}\propto T^{4}\propto a^{-4}$, so Eq.~(\ref{friedman})
gives the familiar relation $H=\left( 2t\right) ^{-1}$. The
temperature descends like $t^{-1/2}$, and at $T=T_{c}$ the phase
transition begins, since we are assuming that no supercooling
occurs. Hence we can use the relation $H_{i} =1/2t_{i}$ as an
initial condition. The scale factor thus takes the very simple
form
\begin{equation}
\left( \frac{a}{a_{i}}\right) ^{3}=4\sin ^{2}\left(
\frac{\sqrt{3}}{4}\frac{t-t_{i}}{t_{i}}+\frac{\pi }{6}\right) ,
\label{scalelneg}
\end{equation}
which does not depend on any parameter, i.e., during the
transition the expansion of the Universe seems to be affected
always in the same way, regardless the thermodynamical parameters
of the theory. However, thermodynamics affects the duration of the
phase transition and the subsequent evolution of the Universe, as
we shall see immediately.

In the above equations it is apparent that the dynamics of the phase transition
depends on thermodynamics only through the ratio
\begin{equation}
r\equiv\frac{s_{u}-s_{b}}{s_{u}}=\frac{L}{T_{c}s_{u}}.  \label{r}
\end{equation}
The fraction of volume is thus $f_{b}=r^{-1}\left[ 1-\left( a/a_{i}\right)
^{-3}\right] $, with $a/a_i$ given by Eq. (\ref{scalelneg}). The phase transition
concludes when $f_{b}=1$ [equivalently \cite{s82}, when $s$ in Eq. (\ref{entropy1})
equals $s_{b}$], so its duration can be determined easily. Making use of some
trigonometric algebra,
\begin{equation}
\frac{t_{f}-t_{i}}{t_{i}}=\frac{4}{\sqrt{3}}\arcsin
\frac{\sqrt{3}\left( 1- \sqrt{1-4r/3}\right) }{4\sqrt{1-r}}.
\label{duracionneg}
\end{equation}
Notice that Eq.~(\ref{duracionneg}) fails to give an answer for $r>3/4$. The problem
is that the scale factor given by Eq. (\ref{scalelneg}) reaches a maximum when the
argument of the sinus is $\pi/2$. If $r$ is small enough, that never occurs between
$t_i$ and $t_f$. For $r=3/4$ it occurs at $t=t_{f}$, and for larger $r$ it happens
before $t_f$, which means that the Universe begins to collapse before the phase
transition has completed. This is not surprising. Indeed, if the energy density of the
unbroken phase is given by Eq.~(\ref{rourad}), then the energy density of the broken
symmetry phase is $\rho _{b}\left( T_{c}\right) =\pi ^{2}g_{\ast }T^{4}/30-L$. So,
there is a negative cosmological constant, which will begin to dominate sooner or
later. If $L<\rho _{u}$ (i.e., $r<3/4$), this won't happen during the phase
transition, but below the critical temperature the Universe will collapse.

\begin{figure}[tbh]
\psfrag{ti}{ $t_i$} \psfrag{tf}{ $t_f$} \psfrag{t0}[][]{ $t_0$}
\psfrag{H0}[][r]{$0$ } \psfrag{Hi}{$H_i$} \psfrag{Hf}{$H_f$}
\centering \epsfysize=5cm \leavevmode \epsfbox{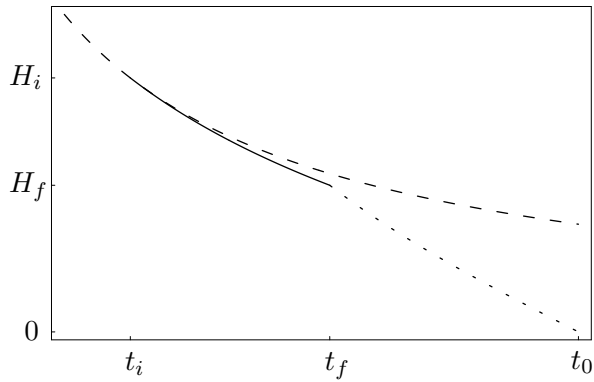}
\caption{The expansion rate of the Universe for a phase transition
at constant $T=T_c$, in the case $\rho _{u} =\pi ^{2}g_{\ast
}T^{4}/30$ (negative cosmological constant).} \label{figequilneg}
\end{figure}

In Fig. \ref{figequilneg} we have plotted the expansion rate of the Universe from the
beginning of the phase transition to the moment at which $H$ becomes zero. The
evolution of $H$ without a phase transition is represented with a dashed line (long
dashes). We have chosen a relatively large value $r=0.5$ in order to get a visible
departure during the phase transition, which occurs between $t_i$ and $t_f$ (solid
line). After the phase transition (short dashes) the expansion slows down, and the
rate eventually vanishes at a time $t_0$.

To consider a more realistic situation we must add a constant term $\sim L$ to the
initial energy density, so that we do not have a cosmological constant of the order of
the scale of the phase transition. We take for simplicity
\begin{equation}
\rho _{u}\left( T\right) =\pi ^{2}g_{\ast }T^{4}/30+L.  \label{roulpos}
\end{equation}
It will be more convenient to re-express the general solution (\ref{scale}) in terms
of the conditions at $t=t_f$, $\left( a/a_{f}\right) ^{3}=\left(
T_{c}s_{b}/p_{c}\right) \sin ^{2}\left[ \omega \left( t-t_{f}\right) +\delta ^{\prime
}\right] $, with $\delta ^{\prime }=\arcsin \sqrt{p_{c}/T_{c}s_{b}}$. From Eq
(\ref{roulpos}) it follows that $\rho _{b}\left( T_{c}\right) =\pi ^{2}g_{\ast
}T_{c}^{4}/30$,  $T_{c}s_{u}=4\rho _{b}/3$, $T_{c}s_{b}=T_{c}s_{u}-L$, and $p_{c}=\rho
_{b}/3-L$. Therefore we can write
\begin{equation}
\left( \frac{a}{a_{f}}\right) ^{3}=\frac{4-4r}{1-4r}\sin
^{2}\left[
\frac{\sqrt{3}\sqrt{1-4r}}{4}\frac{t-t_{f}}{\tilde{t}}+\delta
^{\prime }\right] \label{scalepos}
\end{equation}
with
\begin{equation}
\delta ^{\prime }=\arcsin \sqrt{\frac{1-4r}{4-4r}},
\end{equation}
where we have defined the time scale $\tilde{t}=\left(
2H_{f}\right) ^{-1}$. Before the phase transition, $\rho $ is
given by Eq.~(\ref{roulpos}), so $ a\propto \sinh ^{1/2}\left(
\sqrt{32\pi GL/3}t\right) $. This has the form $a\sim t^{1/2}$ for
$t\ll \left( 32\pi GL/3\right) ^{-1/2}$, and departs from the
radiation-domination behavior unless $\pi ^{2}g_{\ast }T^{4}/30\gg
L$.

The fraction of volume in the broken symmetry phase is
\begin{equation}
f_{b}=\frac{1}{r}\left[ 1-\left( 1-r\right) \left(\frac{ a}{a_{f}}\right) ^{-3}\right]
, \label{fbeqpos}
\end{equation}
and the duration of the phase transition is given by
\begin{equation}
\frac{t_{f}-t_{i}}{\tilde{t}}=\frac{4/\sqrt{3}}{\sqrt{1-4r}}\arcsin
\frac{\sqrt{1-4r}\left( \sqrt{1+4r/3}-1\right) }{\left(4/\sqrt{3}\right)\sqrt{1-r}}.
\label{duracionposrpeq}
\end{equation}
It can be easily checked that, for small $r$, this solution coincides with the
previous one. Furthermore, for $r\rightarrow 0$, the duration of the phase transition
vanishes, as expected. At first sight, there seems to be a problem if $L\geq\rho
_{b}/3$ (i.e., for $r\geq1/4$). However, all the previous expressions are still valid
and real in the range $1/4\leq r<1$. They can be written in the form
\begin{equation}
\left( \frac{a}{a_{f}}\right) ^{3}=\frac{4-4r}{4r-1}\sinh
^{2}\left[
\frac{\sqrt{3}\sqrt{4r-1}}{4}\frac{t-t_{f}}{\tilde{t}}+\delta
^{\prime }\right] ,
\end{equation}
with
\begin{equation}
\delta ^{\prime }=\textrm{arcsinh}\sqrt{\frac{1-4r}{4-4r}},
\end{equation}
and
\begin{equation}
\frac{t_{f}-t_{i}}{\tilde{t}}=\frac{4/\sqrt{3}}{\sqrt{4r-1}} \textrm{arcsinh}
\frac{\sqrt{4r-1}\left( \sqrt{1+4r/3}-1\right) }{\left(4/\sqrt{3}\right)\sqrt{1-r}}.
\end{equation}
For $r\rightarrow 1$ the duration of the phase transition becomes infinite because the
constant energy density $L$ is comparable to the energy density of the radiation,
playing the role of a cosmological constant that starts dominating at $T\sim T_{c}$.
Therefore the expansion of the Universe becomes too fast and the phase transition
never ends. One expects that some of our initial assumptions will break down near this
limit. For instance, the temperature and expansion rate will not be homogeneous, due
to the significant energy density contrast between the two phases and the rapid
expansion of the Universe.

\begin{figure}[tbh]
\psfrag{ti}{ $t_i$} \psfrag{tf}{ $t_f$} \psfrag{Hi}{$H_i$}
\psfrag{Hf}{$H_f$} \centering \epsfysize=5cm \leavevmode
\epsfbox{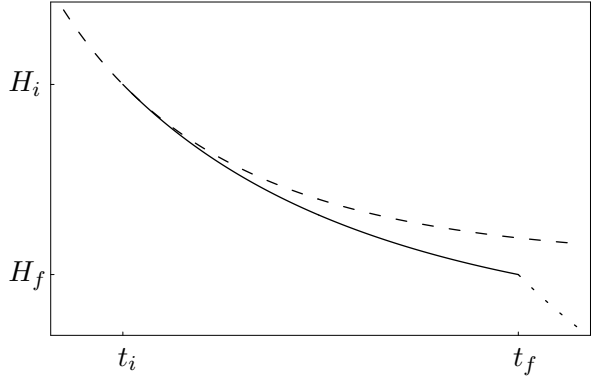} \caption{The expansion rate of the Universe
for a phase transition at constant $T=T_c$, in the case $\rho _{u}
=\pi ^{2}g_{\ast }T^{4}/30+L$.} \label{figequilpos}
\end{figure}

In Fig.~\ref{figequilpos} we plot the expansion rate as a function of time for
$r=0.8$. We see that without the phase transition the Universe would enter exponential
expansion ($H\rightarrow $ constant). After the phase transition the evolution returns
to the radiation-dominated relation $H= 1/2t$.  By the end of the transition the
departure of the expansion rate from its previous evolution becomes appreciable,
because in this case $t_f-t_i\sim \tilde{t}\sim t_i$. If the duration of the phase
transition is short in comparison with the age of the Universe,  the back-reaction on
$H$ can be disregarded. According to Eq.~(\ref{duracionposrpeq}), this happens when
the energy released is small in comparison to the energy density of the plasma (small
$r$).

\section{Supercooling \label{superc}}

In the case of a phase transition mediated by homogeneous
nucleation, bubbles start to nucleate at a temperature $T<T_{c}$,
when the gain in free energy inside a bubble is enough to
compensate the cost of gradient energy at the surface. We have
seen in the previous section that, even if bubbles begin to grow
at $T=T_{c}$, the phase transition may not come to an end if the
parameter $r$ is close to $1$. The case of homogeneous nucleation
is even worse due to the additional supercooling. As we shall see,
a large latent heat is a general feature of strongly first-order
phase transitions. In this case there may be extreme supercooling
from which the Universe may never recover \cite{gw81,gw83}.
Considerable supercooling and latent heat release may occur for
instance in the quark-hadron phase transition \cite{qcd}. Notice
that in the case of large supercooling there will be an important
departure from equilibrium, at least at the beginning of the phase
transition. For the rest of the paper we will be mostly interested
in the case $r\ll 1$.

The nucleation and growth of bubbles in a first order phase
transition has been extensively studied in the context of the QCD
\cite{qcd,hkllm93} and electroweak
\cite{h95,ah92,dlhll92,eikr92,e96,hkllm93} phase transitions.
After a bubble is formed, it grows due to the pressure difference
at its surface. There is a very short acceleration stage until the
wall reaches a terminal velocity due to the friction of the
plasma. It can be seen that this initial period in the history of
the bubble expansion is negligible. We will assume again that the
system remains close to equilibrium, in accordance with the
assumption that the velocity of the bubble wall is small. If the
wall velocity is less than the speed of sound in the relativistic
plasma, $v_{w}<c_{s}=\sqrt{1/3}$, the wall propagates as a
deflagration front. This means that a shock front precedes the
wall, with a velocity $v_{sh}>c_{s}$. For $v_{w}\ll c_{s}$, the
latent heat is transmitted away from the wall and quickly
distributed throughout space. We can take into account this effect
by considering a homogeneous reheating of the plasma during the
expansion of bubbles \cite{h95,m01}. (For detailed treatments of
hydrodynamics at different wall velocities see, e.g.,
\cite{eikr92,hkllm93}).

\subsection{Bubble nucleation}

The thermal tunnelling probability for bubble nucleation per unit
volume and time is \cite{a81,l83}
\begin{equation}
\Gamma \simeq A\left( T\right) e^{-S_{3}/T}.  \label{gamma}
\end{equation}
The prefactor involves a ratio of determinants associated with the
quantum fluctuations around the instanton. In general it must be
evaluated numerically. It is usually assumed to be roughly of
order $T^{4}$, since the nucleation rate is dominated by the
exponential in (\ref{gamma}). We will consequently assume $
A\left( T\right) \simeq T_{c}^{4}$. $S_{3}\left( T\right) $ is the
three-dimensional instanton action, which coincides with the free
energy of a critical bubble (i.e., a bubble in unstable
equilibrium between expansion and contraction),
\begin{equation}
S_{3}=4\pi \int_{0}^{\infty }r^{2}dr\left[ \frac{1}{2}\left(
\frac{d\phi }{dr }\right) ^{2}+V\left( \phi \left( r\right)
,T\right) \right] .  \label{s3}
\end{equation}

The configuration of the nucleated bubble may be obtained by
extremizing this action. Hence it obeys the equation
\begin{equation}
\frac{d^{2}\phi }{dr^{2}}+\frac{2}{r}\frac{d\phi }{dr}=V^{\prime }\left(
\phi \right) .  \label{eqprofile}
\end{equation}
For temperatures very close to $T_{c}$, the width of the bubble wall at the moment of
formation is much smaller than its radius, and a thin wall approximation can be used
\cite{l83,c77}, in which $S_{3}$ is expressed as a function of the critical bubble
radius $R_{c}$, the free energy difference $ V $ between the two minima of the
potential, and the bubble wall surface energy $\sigma $. The radius $R_{c}$ can thus
be obtained by finding the maximum of $S_{3}$. A similar approximation can be used to
estimate the free energy and radius of a thick-walled bubble when temperature is not
so close to $T_{c}$ \cite{ah92}. However, as pointed out in Ref.~\cite{dlhll92}, due
to the exponential dependence, the tunnelling probability may be strongly
overestimated by using approximations to $S_{3}$, leading to a sooner completion of
the phase transition\footnote{In many cases the phase transition occurs in a tiny
range of temperature about $T_{c}$, so it is a good approximation to replace almost
every quantity by its value at $T\simeq T_{c}$. The important exception are quantities
such as $S_{3}$, that depend directly on the free energy difference $V\left( \phi
,T\right) $, which varies drastically with $T$ at the critical temperature
\cite{m00}.}. Since we do not intend to do numerical calculations in the present work,
we will use the thin wall approximation. This approximation may be reasonable or not,
depending on the amount of supercooling.

\subsection{Phase transition dynamics}

In the previous section we assumed constant temperature, and used
Eqs.~(\ref{entropy1}) and (\ref{entropy2}) to obtain the fraction
of volume $f_{b}$ in terms of the scale factor $a$. Then the
Friedman equation determined $a\left( t\right) $. In the present
case the temperature is not constant, so we need an extra equation
to solve for the three quantities $ f_{b}$, $a$ and $T$. Such an
equation arises by considering the nucleation and growth of
bubbles \cite{gw81},
\begin{equation}
f_{b}\left( t\right) =1-\exp \left\{ -\frac{4\pi }{3}\int_{t_{i}}^{t}\left[
\frac{a\left( t^{\prime }\right) }{a\left( t\right) }\right] ^{3}\Gamma
\left( T^{\prime }\right) R\left( t^{\prime },t\right) ^{3}dt^{\prime
}\right\} ,  \label{fb}
\end{equation}
where $T^{\prime }$ is the temperature at $t=t^{\prime } $,
$t_{i}$ is the time at which the Universe reaches the critical
temperature,
\begin{equation}
t_{i}\simeq \xi M_{P}/T_{c}^{2},
\end{equation}
where $\xi =\sqrt{90/32\pi ^{3}g_{\ast }}$, and $R\left( t^{\prime
},t\right) $ is the radius of a bubble that nucleated at time
$t^{\prime }$ and expanded until the time $t$,
\begin{equation}
R\left( t^{\prime },t\right) =R_{c}\left( T^{\prime }\right) \frac{a\left(
t\right) }{a\left( t^{\prime }\right) }+\int_{t^{\prime }}^{t}v_{w}\left(
T^{\prime \prime }\right) \frac{a\left( t\right) }{a\left( t^{\prime \prime
}\right) }dt^{\prime \prime }.  \label{radius}
\end{equation}
The factors of $a$ in Eqs.~(\ref{fb}) and (\ref{radius}) take into
account the fact that the number density of nucleated bubbles is
diluted, and the radius of a bubble enlarged, due to the expansion
of the Universe from $t^{\prime }$ to $t$ (see e.g. \cite{bcv00}).
We can assume that this effect is negligible if the duration of
the phase transition is small in comparison with the age of the
Universe. As we have seen in the previous section, this is true
when $ L/\rho _{b}\ll 1$, which is the case we will consider.
(This assumption will not hold, in principle, for the QCD phase
transition). The wall velocity $v_{w}$ is determined by the
equilibrium between the pressure difference $V\left( T\right) $
and the friction force exerted by the plasma. The latter is
proportional to the wall velocity. The constant of proportionality
is the friction coefficient $\eta $ (see section \ref{physquant}),
so
\begin{equation}
v_{w}=-V\left( T\right) /\eta .  \label{vw}
\end{equation}
 The exponent in
Eq.~(\ref{fb}) is minus the fraction of volume occupied by bubbles
that nucleated between $t_{i}$ and $t$, if we do not take into
account overlapping of bubbles. At the beginning of nucleation the
formula $ f_{b}\left( t\right) \simeq (4\pi /3)\int \Gamma \left(
T^{\prime }\right) R\left( t^{\prime },t\right) ^{3}dt^{\prime }$
is correct.

Using again Eqs.~(\ref{entropy1}) and (\ref{entropy2}), we obtain
the analogous of Eq.~(\ref{fractiontc}),
\begin{equation}
f_{b}=\frac{1}{s_{u}\left( T\right) -s_{b}\left( T\right) }\left(
s_{u}\left( T\right) -s_{u}\left( T_{c}\right)
\frac{a_{i}^{3}}{a^{3}} \right) ,  \label{fractionsuperc}
\end{equation}
but since we already have an equation for $f_{b}$, namely, Eq.
(\ref{fb}), we use Eq. (\ref{fractionsuperc}) to express $T$ in
terms of $f_{b}$ and $a$,
\begin{equation}
T^{3}=\frac{V^{\prime }\left( T\right) }{2\pi ^{2}g_{\ast
}/45}f_{b}+\frac{T_{c}^{3}a_{i}^{3}}{a^{3}}.  \label{temperature}
\end{equation}
Eq.~(\ref{temperature}) has come across within an approach that
differs from previous works, so it is worthwhile spending a few
words discussing its physical meaning. We may follow for instance
Ref. \cite{h95}, and use energy (non-) conservation in the
following way. On one hand, we may write the total energy in a
volume $V_{U}=V_{u}+V_{b}$ as $E=\left[ \rho _{u}+\Delta \rho
f_{b}\right] V_{U}=\rho V_{U}$, where $\Delta \rho =\rho _{b}-\rho
_{u}$. If the Universe were not expanding, energy conservation
during the phase transition would give
\begin{equation}
\dot{\rho}=\dot{\rho}_{u}+\Delta \dot{\rho}f_{b}+\Delta \rho \dot{f}_{b}=0.
\label{heck1}
\end{equation}
This gives the rate $\dot{\rho}_{u}$ at which the plasma takes
energy from the change of phase. On the other hand, when it is not
undergoing a phase transition, the Universe takes energy from the
plasma at a rate
\begin{equation}
\dot{\rho}_{u}=\dot{\rho}=-4\rho _{u}H.  \label{heck2}
\end{equation}
If we join the two equations, we obtain the total rate of change
of energy as
\begin{equation}
\dot{\rho}_{u}=-\Delta \dot{\rho}f_{b}-\Delta \rho \dot{f}_{b}-4\rho _{u}H,
\label{heckler}
\end{equation}
from where we get an equation for $\dot{T}$. However, if the phase
transition and the expansion of the Universe are taken into
account at the same time, additional terms appear both in
Eqs.~(\ref{heck1}) and (\ref{heck2}). On one hand, there is a term
of the form $\rho \dot{V}_{U}$ in $ \dot{E}$, which produces a new
term $-3H\rho $. This just accounts for energy dilution. On the
other hand, the expansion of the Universe takes energy not only
from radiation, since it is not the only component in the equation
of state. Bearing in mind the two coexisting phases during the
phase transition, the rate at which energy conservation is
violated is given by $dE=-p_{u}dV_{u}-p_{b}dV_{b}$. Using this to
obtain $\dot{\rho}$, we finally find
\begin{equation}
\dot{\rho}_{u}=-\Delta \dot{\rho}f_{b}-\Delta \rho \dot{f}_{b}-3H\left( \rho
_{u}+p_{u}+\Delta \rho f_{b}+\Delta pf_{b}\right) -\Delta p\dot{f}_{b},
\label{noheckler}
\end{equation}
where $\Delta p=p_{b}-p_{u}$. Since $\rho +p=Ts$, it can be seen that the
first terms of this equation reproduce Eq.~(\ref{heckler}), but there are
additional terms. Using $s=dp/dT$ and rearranging Eq.~(\ref{noheckler}), we
see that it is just the equation for entropy conservation,
\begin{equation}
\dot{s}_{u}+\Delta \dot{s}f_{b}+\Delta s\dot{f}_{b}=-3H\left( s_{u}+\Delta
sf_{b}\right) ,  \label{entropycons}
\end{equation}
from which we may re-obtain the result (\ref{temperature}). The discrepancies between
Eq.~(\ref{heckler}) and Eq.~(\ref{noheckler}) will not be important as long as the
latent heat $L$ is not significant and $T$ remains close to $T_{c}$. Indeed, we can
neglect the last two terms inside the parenthesis in Eq.~(\ref{noheckler}), provided
that $\Delta \rho \ll \rho $. The last term in Eq.~(\ref{noheckler}) is responsible
for the appearance of the entropy difference $\Delta s=-V^{\prime }\left( T\right) $
as the factor of $\dot{f}_{b}$ in Eq.~(\ref{entropycons}), instead of the energy
difference $\Delta \rho =V\left( T\right) -TV^{\prime }\left( T\right) $. Since
$V\left( T_{c}\right) =0$, the two quantities are related by $\Delta \rho \simeq
T\Delta s$ for $T\simeq T_{c}$. Therefore, Eq. (\ref{heckler}) gives a good
approximation in the case $r\ll 1$. Still, the fact that in Eq. (\ref{temperature})
the temperature is already integrated may constitute an advantage for analytical
calculations.

Finally, the evolution of the scale factor is given by the
Friedman equation,
\begin{equation}
H^{2}=\frac{8\pi G}{3}\left[ \rho _{u}\left( T\right) +\Delta \rho
f_{b} \right] .  \label{friedmansuperc}
\end{equation}
If $\Delta \rho \ll \rho $, then one can use the customary formula $H\propto
T^{2}$ for the expansion rate. In fact, the variation of $\rho _{u}$ is of
the same order of $\Delta \rho $, so if we neglect $\Delta \rho $, to be
consistent we should also set $H=const\propto T_{c}^{2}$ in this
approximation. Again, this is reasonable if the duration of the phase
transition is short enough.

In each particular model, Eqs.~(\ref{fb}), (\ref{temperature}) and
(\ref{friedmansuperc}) can be used to calculate numerically the
evolution of the phase transition. In this paper, dough, we will
make analytical estimates of the parameters that are relevant for
the cosmological consequences, such as the temperature and wall
velocity in the different stages of the transition.

\subsection{Bubble coalescence}

When bubbles occupy more than 30\% of space, they meet and
percolate. This occurs when bubbles have a characteristic radius
$R_{0}\sim \left( 0.3/n_{b}\right) ^{1/3}$, where $n_{b}$ is the
number density of bubbles. Bubble coalescence provides a different
mechanism of bubble growth, in which the driving force is surface
tension instead of pressure difference. When bubbles collide and
percolate, they arrange themselves into a system of fewer, larger
bubbles in order to minimize the surface area. We can estimate the
characteristic time of this process as follows \cite{w84}. Assume
that when two spherical bubbles of radius $R$ meet, they form a
single bubble of radius $2^{1/3}R$. In this process, a mass of
fluid $m\sim \rho R^{3}$ is moved a distance of order $R$. If this
is done in a time $t$, the kinetic energy involved in the process
is $K\sim m\left( R/t\right) ^{2}$. This energy is supplied by the
surface of the bubbles. The surface energy released in the process
is $\Delta E\sim \sigma R^{2}$, so this occurs in a time $t\sim
\left( \rho R^{3}/\sigma \right) ^{1/2}$.

Once bubble coalescence begins, it will be the fastest mechanism
of bubble growth if the rate $t^{-1}$ is larger than the rate
$v_{w}/R$ given by the wall velocity (\ref{vw}). Therefore this
process will dominate until bubbles reach a characteristic size
$R_{1}\sim \sigma /v_{w}^{2}\rho $. If $ R_{1}<R_{0}$, then it
never dominates. Otherwise, during the period in which the radius
varies from $R_{0}$ to $R_{1}$ bubble coalescence is the fastest
process and must be taken into account.

The process of bubble coalescence may end in two different ways.
If $v_{w}$ is large enough, the radius $R_{1}$ is very close to
$R_{0}$, and after a short period bubbles continue to grow with
velocity $v_{w}$. If, on the contrary, $v_{w}$ is very small, then
coalescence dominates for a larger period, until the
low-temperature phase occupies more than 50\% of the total volume.
At this moment the regions of high-temperature phase detach into
isolated bubbles and the process stops. The interface velocity is
again determined by pressure difference, friction, and latent heat
release, so these bubbles shrink with velocity $v_{w}$. This
occurs at a bubble radius $ R_{2}\sim \left( 0.5/n_{b}\right)
^{1/3}$. Notice that, although this expression is similar to that
for $R_{0}$, in fact $R_{2}$ may be very different from $R_{0}$
since $n_{b}$ may decrease significantly during the process due to
the expansion of the Universe.

Finally, when bubbles of the high temperature phase are so small
that surface energy becomes dominant over volume energy, the
shrinking is accelerated until the symmetric-phase bubbles
disappear or, eventually, until a topological defect is formed.

\section{Bubble number and interface velocity \label{transicion}}

In the case of homogeneous nucleation, the Universe cools down to
a temperature $T_{N}<T_{c}$ before bubble nucleation becomes
appreciable, so the phase transition effectively starts at a time
$t_{N}>t_{i}$. Then, the phase transition proceeds in two main
steps. At the beginning bubbles nucleate with a rate $\Gamma
\left( T_{N}\right) $, and expand with a velocity given by the
friction coefficient $\eta $ and the pressure difference $V\left(
T_{N}\right) $. After a short time the energy released by the
change of phase reheats the Universe up to a temperature $T_{r}$
close to $T_{c}$. Then, a longer period begins, in which the two
phases are close to equilibrium. How long is this period, and how
close is $T_r$ to $T_c$, depends on the latent heat. The free
energy difference in this stage is $V\left( T_{r}\right) \simeq
0$, so the velocity of bubble expansion decreases and the bubble
nucleation rate becomes extremely suppressed. Therefore, this
second stage may be very similar to the inhomogeneous nucleation
phase transition discussed in Section \ref{phaseeq}.

In order to avoid numerical calculations, we will need some
approximations for the fraction of volume $f_{b}$, the temperature
$T$, and the expansion rate of the Universe $H$ [formulas
(\ref{fb}), (\ref{temperature}) and (\ref{friedmansuperc})]. We
notice that the latter cannot be affected by the phase transition
until the energy released becomes appreciable. Furthermore, when
this occurs, the plasma has reheated up to $T_{r}\simeq T_{c}$,
and enters the phase equilibrium stage. We have seen in Section
\ref{phaseeq} that $H$ is not modified significantly if the
parameter $r$ defined in (\ref{r}) is small. So, we do not need to
consider back-reaction on the scale factor and we will assume that
the evolution of the Universe does not depart from the standard
relations $H\simeq 1/2t$, $a\propto t^{1/2}$. We will also need an
approximation for the nucleation rate and free energy difference.

\subsection{Thin wall approximation and linearization of $V$}

If the width of the wall is much less than the radius of the
bubble, we can neglect the second term in Eq.~(\ref{eqprofile}) to
obtain the wall profile (this approximation is exact at the
critical temperature, at which $ R_{c}\rightarrow \infty $, and
gives the usual  kink profile). If we then multiply by $d\phi /dr$
and integrate using the boundary conditions $d\phi /dr=0$ and
$V=0$ outside the bubble, we find that
\begin{equation}
\frac{1}{2}\left( \frac{d\phi }{dr}\right) ^{2}=V\left( \phi \right) ,
\label{thinwallapp}
\end{equation}
so, $d\phi /dr=-\sqrt{2V}$ since at the wall $\phi $ falls from
$\phi _{m}$ to $0$. Inserting this in Eq.~(\ref{s3}) we obtain the
free energy of the critical bubble in the thin wall approximation,
\begin{equation}
S_{3}=\frac{4\pi }{3}R_{c}^{3}V\left( T\right) +4\pi R_{c}^{2}\sigma \left(
T\right) ,
\end{equation}
where the  free energy difference $V\left( T\right) $ is defined in Eq.
(\ref{deltav}), and $\sigma \left( T\right) $ is the surface tension of the bubble
wall,
\begin{equation}
\sigma =\int \left( \frac{d\phi }{dr}\right) ^{2}dr=\int_{0}^{\phi
_{m}} \sqrt{2V}d\phi .  \label{formulasigma}
\end{equation}
Maximizing with respect to $R_{c}$ we get the values of the
critical radius and action,
\begin{eqnarray*}
R_{c} &=&-2\sigma /V, \\
S_{3} &=&16\pi \sigma ^{3}/3V^{2}.
\end{eqnarray*}
Since $\sigma $ does not change significantly during the phase transition,
it can be approximated by $\sigma \left( T_{c}\right) $.

Since the thin wall approximation is valid when $T$ does not depart significantly from
$T_{c}$, we can also make a linear approximation for the free energy difference,
\begin{equation}
V\left( T\right) \simeq L\left( T-T_{c}\right) /T_{c}.  \label{linearv}
\end{equation}
So, the exponent in the nucleation rate (\ref{gamma}) becomes
\begin{equation}
\frac{S_{3}\left( T\right) }{T}\simeq\frac{16\pi \sigma ^{3}T_{c}}{3L^{2}\left(
T_{c}-T\right) ^{2}},  \label{s3thin}
\end{equation}
and the critical radius is
\begin{equation}
R_{c}\simeq 2\sigma T_{c}/L\left( T_{c}-T\right) .  \label{rc}
\end{equation}
With the linear approximation (\ref{linearv}), the wall velocity is given by
\begin{equation}
v_{w}=L\left( T_{c}-T\right) /\eta T_{c}.  \label{vwlinear}
\end{equation}

\subsection{The first stage: nucleation and reheating}

Neglecting the effects of latent heat, the phase transition may be assumed to occur
roughly at $t=t_{N}$, since it goes on for a very short interval $\delta t\ll t_{N}$
\cite{m00}. With the inclusion of latent heat the phase transition evolves during a
longer period, that begins at $t_{N}$. In this case, we expect that the first part of
the evolution, in which the Universe is reheated up to a temperature $T_{r}$, has a
time scale of the same order of the time interval $\delta t$ of the phase transition
without latent heat. This is confirmed by numerical calculations \cite{m01}. The
nucleation rate $\Gamma $ vanishes at $T=T_{c}$, but it changes very quickly with
temperature and becomes of order $T^{4}$ at $T=T_{0}$, where the barrier between the
two minima of the free energy disappears. This is an extremely large rate, so it is
impossible that the Universe supercools close to $T_{0}$ \cite{m00}. We will assume
that the temperature $T_N$ is close enough to $T_c$ that the approximations of the
previous subsection can be used.

In Ref.~\cite{ah92} the onset of nucleation was assumed to occur
when the probability that a bubble was nucleated inside each
causal volume is one,
\begin{equation}
\int_{t_{i}}^{t_{N}}V_{H}\Gamma dt\sim 1.
\end{equation}
The causal volume is given by $V_{H}=d_{H}^{3}$, where the horizon size $ d_{H}$
scales like the age of the Universe, $d_{H}\sim 2t$. The cosmological scale $t$,
however, is in general too large in comparison with the scale of phase transition
dynamics, which at $t=t_N$ is given by $t_N-t_i$. The scale of phase transition
dynamics is roughly determined by the temperature variation during the phase
transition, which is bounded by the difference $T_{c}-T_{0}$.

Consequently, it may be more appropriate to consider a different causal distance
$d_{c}$, related to the dynamics of the phase transition in the following way. We may
say that bubbles begin to \textquotedblleft see\textquotedblright\ each other at a
time $t_{N}$ when their mean separation is of the order of the distance travelled by a
sound wave since time $t_{i}$. Then, the causal distance is given by $d_{c}\sim
c_{s}\left( t_{N}-t_{i}\right) $, where $ c_{s}=1/\sqrt{3}$ is the velocity of sound
in the relativistic fluid. This defines a causal volume in terms of $T_{N}$
\begin{equation}
V_{c}=\left[ c_{s}\xi M_{P}\left( 1/T_{N}^{2}-1/T_{c}^{2}\right) \right]
^{3},  \label{causal}
\end{equation}
We remark that the real improvement in using Eq.~(\ref{causal}) instead of $ V_{H}$
does not come from considering the velocity of sound, which is $\sim 1 $, but from the
fact that in many cases the time elapsed from $t_{i}$ to $ t_{N}$ will be much less
than the age of the Universe $t_{i}\simeq \xi M_{P}/T_{c}^{2}$. The volume $V_{c}$ is
thus suppressed with respect to $ V_{H}$ by a factor $\left[ \left( T_{i}-T_{N}\right)
/T\right] ^{3}$. The nucleation time $t_{N}$ calculated in this way is larger, since
more bubbles need to be nucleated before they are separated by a distance at which
they are causally connected to each other.

There will be a bubble in each volume $V_{c}$ when
\begin{equation}
\int_{t_{i}}^{t_{N}}V_{c}T^{4}e^{-S_{3}\left( T\right) /T}dt\sim 1.
\label{eqonset}
\end{equation}
To evaluate the integral in (\ref{eqonset}) we make use of the following approximation
\cite{ah92} (see also \cite{eikr92}). The three-dimensional action in Eq.
(\ref{s3thin}) can be expanded about any temperature $T_{\ast }$ in the form
\begin{equation}
\frac{S_{3}\left( T\right) }{T}=\frac{S_{3}\left( T_{\ast }\right)
}{T_{\ast }}\frac{1}{\left( 1-x\right) ^{2}}=\frac{S_{3}\left(
T_{\ast }\right) }{ T_{\ast }}\left( 1+2x+\cdots \right) ,
\label{s3expanded}
\end{equation}
where $x=\left( T-T_{\ast }\right) /\left( T_{c}-T_{\ast }\right) $. Since the
integrand in Eq.~(\ref{eqonset}) is sharply peaked at $T_{N}$, we choose $ T_{\ast
}=T_{N}$ and use the expansion (\ref{s3expanded}) to evaluate the integral. We find
\begin{equation}
S_{3}\left( T_{N}\right) /T_{N}\simeq 4\log \frac{2\xi M_{P}}{T_{N}}+4\log
\frac{T_{c}-T_{N}}{T_{c}}-\log \frac{S_{3}\left( T_{N}\right) }{T_{N}},
\end{equation}
which will be in general dominated by the first term. For
temperatures $T$ several orders of magnitude below $M_{P}$, we
have $S_{3}\left( T_{N}\right) /T_{N}\gtrsim 100$ (it is e.g.
$\sim 140$ for the electroweak scale, and $ \sim 180$ for the QCD
scale). From Eq.~(\ref{s3thin}) we obtain
\begin{equation}
\left( \frac{T_{c}}{T_{c}-T_{N}}\right) ^{2}\simeq
\frac{3L^{2}T_{c}}{16\pi \sigma ^{3}}\left( 4\log \frac{2\xi
M_{P}}{T_{c}}+\log \frac{3L^{2}T_{c}}{ 8\pi \sigma ^{3}}+6\log
\frac{T_{c}-T_{N}}{T_{c}}\right) ,  \label{ti}
\end{equation}
where we have used that $T_{N}\simeq T_{c}$. In some of the following estimations we
will replace $T_{c}-T_{N}$ by $T_{c}-T_{0}$ inside the $\log $'s, since both
differences are generally of the same order.

Immediately after $t=t_{N}$, the temperature increases at a rate which is given by Eq.
(\ref{noheckler}). Under the current approximations we can write  $\dot{\rho}=L\dot{f}
_{b}-4\rho H$. Therefore, the rate of change of energy of the plasma is
$\dot{\rho}/\rho \simeq (3r/4)\dot{f}_{b}-4H$. Thus, on one hand energy is increased
at a rate $\sim r\delta f_{1}/\delta t_{1}$, where $\delta f_{1}$ is the fraction of
volume converted to the low-T phase during the reheating stage, and $\delta t_{1}$ is
the duration of this stage. The rate of energy decrease, on the other hand, is given
by $H\sim 1/t$. Since $\delta t_{1}$ is much shorter than the age of the Universe, the
increasing rate is much larger than the latter. The total change in energy density is
thus $\delta \rho \sim L\delta f_{1}$.

The temperature cannot increase beyond $T_{c}$, so if $L\gtrsim \rho \left(
T_{c}\right) -\rho \left( T_{N}\right) $, the fraction $\delta f_{1}$ will be less
than one. This means that before the phase transition completes, a final temperature
$T_{r}$ very close to $T_{c}$ is reached, and the phase transition proceeds more
slowly until $\delta f =1$. For $L< \rho \left( T_{c}\right) -\rho \left(
T_{N}\right)$ we have a variation $\delta \rho \sim L$ during the phase transition,
which gives us an idea of the reheating that occurs. There may be significant
reheating and a considerable variation of the wall velocity, but there will not be a
long phase equilibrium stage, since in this case $\delta f_{1}\simeq 1$.  The case
$L\ll \rho \left( T_{c}\right) -\rho \left( T_{N}\right) $ corresponds to weakly
first-order phase transitions, which most likely occur by spinodal decomposition
rather than by nucleation and expansion of bubbles \cite{g94,m00}. We could also have
$L\gg \rho \left( T_{c}\right) -\rho \left( T_{N}\right) $, although we are assuming
$L\ll \rho \left( T_{c}\right) $. In this case we have $\delta f_{1}\ll 1$, which
means that most of the phase transition happens with the two phases near equilibrium,
and we can apply the analysis of Section \ref{phaseeq}. In this section we thus
concentrate in the case in which the latent heat is comparable with the energy
difference $ \rho \left( T_{c}\right) -\rho \left( T_{N}\right) $.

The expansion of bubbles is governed by Eq.~(\ref{radius}). At $T=T_{N}$ bubbles
nucleate with a radius $R_{c}$ given by Eq.~(\ref{rc}), and after a time $\delta t$
the radius has increased an amount $v_{w}\delta t$, with $ v_{w}$ given by
Eq.~(\ref{vwlinear}). The ratio of the two distances is
\begin{equation}
\frac{v_{w}\delta t}{R_{c}}\simeq \frac{L^{2}\left( T_{c}-T_{N}\right)
^{2}\delta t}{2\eta \sigma T_{c}^{2}}.
\end{equation}
For the time scale of the phase transition dynamics, $\delta t\sim
\xi M_{P}\left( T_{c}-T_{0}\right) /T^{3}$, we have
\begin{equation}
\frac{v_{w}\delta t}{R_{c}}\sim \frac{L^{2}\xi }{\eta \sigma
T}\frac{M_{P}}{T }\left( \frac{T_{c}-T_{0}}{T}\right) ^{3}.
\end{equation}
Thus, if the phase transition takes place at a temperature sufficiently
below the plank scale, the bubbles will grow so rapidly that we can safely
neglect the initial radius $R_{c}$ in Eq.~(\ref{radius}).

\begin{figure}[tbh]
\psfrag{ti}[][b]{ $t_i$} \psfrag{tf}[][b]{ $t_f$}
\psfrag{tm}[][b]{ $t_m$} \psfrag{pp}[][l]{$T_c$}
\psfrag{Tm}[][l]{$T_m$} \psfrag{dt1}[][b]{$\delta t_1$}
\psfrag{dt2}[][b]{$\delta t_2$} \centering \epsfysize=5.5cm
\leavevmode \epsfbox{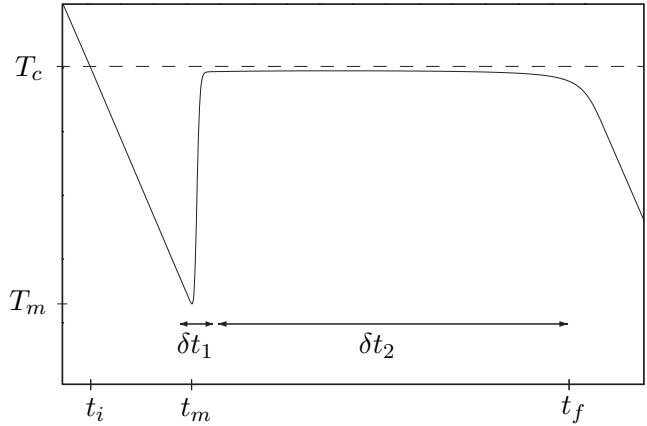} \caption{Typical evolution of
the temperature during a phase transition with supercooling.}
\label{esquema}
\end{figure}

At the beginning, bubbles expand with constant velocity
$v_{w}\left( T_{N}\right) $. When reheating becomes important, the
bubble expansion slows down and we enter the second stage.
Consequently, during the first stage $ \dot{f}_{b}$ can be roughly
estimated without taking into account the liberation of latent
heat. We can thus estimate the rate $\dot{T}/T\sim r \dot{f}_{b}$
without considering the back-reaction of the reheating on
$\dot{f}_{b}$. It is only at the beginning and at the end of
reheating that $r\dot{f}_{b}\sim H$ and $\dot{T}\simeq 0$. Indeed,
soon after $t=t_{N}$, the temperature takes its minimum value
$T_{m}\lesssim T_{N}$ (See Fig.~\ref{esquema}), then it increases
to $T_{r}\lesssim T_c$. The value of $T_m$ is important because
the nucleation rate turns on at $t\simeq t_{N}$, is maximal at $
t=t_{m}$, and turns off again when the temperature has increased
back to $ T_{N}$. All the process occurs in a time $\delta
t_{\Gamma }\sim \left( t_{m}-t_{N}\right) $, which is determined
by the speed at which temperature changes. This interval $\delta
t_{\Gamma }$ in which $\Gamma $ is not negligible is much less
than $\delta t_1$, due to the exponential dependence of $\Gamma $
on $T_{c}-T$.

Accordingly, the temperature $ T_{m}$ occurs when
\begin{equation}
r4\pi v_{w}\left( T_m\right)\int_{t_{i}}^{t_{m}}\Gamma \left( T\right) v_{w}^{2}\left(
T\right) \left( t_{m}-t\right) ^{2}dt\simeq H.
\end{equation}
where the time-temperature relation $t=\xi M_{P}/T^{2}$ must be used to evaluate the
integral. Using the expansion (\ref{s3expanded}) about $T_{\ast }=T_{m}$,  we find
\begin{equation}
\left[ \frac{S_{3}\left( T_{m}\right) }{T_{m}}\right]
^{-1}e^{-\frac{ S_{3}\left( T_{m}\right) }{T_{m}}}\simeq
\frac{2\pi \sigma ^{6}\eta ^{3}T_{c}^{2}}{9\xi ^{6}L^{8}}\left(
\frac{T_{c}}{M_{P}}\right) ^{4}\left(
\frac{T_{c}}{T_{c}-T_{m}}\right) ^{10}.  \label{sm}
\end{equation}

The number density of bubbles is given by
\begin{equation}
n_{b}=\int_{t_{c}}^{t_{f}}\Gamma \left( t\right) dt.
\end{equation}
Since $\Gamma \left( t\right) $ is sharply peaked at $t_{m}$, we can estimate $n_b$ as
$2\int_{t_{c}}^{t_{m}}\Gamma \left( t\right) dt$ and use again the dependence
$t\propto T^{-2}$. The integral can be calculated again using the expansion
(\ref{s3expanded}), which gives
\begin{equation}
n_{b}\sim T_{c}^{3}\frac{\xi M_{P}}{T_{c}}\frac{T_{c}-T_{m}}{T_{c}}\left[
\frac{S_{3}\left( T_{m}\right) }{T_{m}}\right] ^{-1}e^{-S_{3}\left(
T_{m}\right) /T_{m}}.  \label{nb1}
\end{equation}
We may define the interval $\delta t_{\Gamma }$ by writing $n_{b}\simeq
\Gamma \left( t_{m}\right) \delta t_{\Gamma }$. Then we can easily see in
Eq. (\ref{nb1}) that
\begin{equation}
\delta t_{\Gamma }\sim \left[ S_{3}\left( T_{m}\right) /T_{m}\right)] ^{-1}\left(
t_{m}-t_{i}\right) .
\end{equation}
Therefore, $t_{m}-t_{N}$ is much less than $t_{m}-t_{i}$ (most
typically, about two orders of magnitude~\footnote{This shows that
it is more accurate to estimate the difference $t_{m}-t_{N}\sim
\delta t_{\Gamma }$ in this way, rather than subtracting the
values given by Eqs. (\ref{ti}) and (\ref{sm}).}). Using the
result (\ref{sm}), with $T_{m}\simeq T_{N}$, we finally obtain the
density of bubbles,
\begin{equation}
n_{b}\sim \frac{2\pi \sigma ^{6}\eta ^{3}T_{c}^{2}}{9\xi ^{5}L^{8}}\left(
\frac{T_{c}}{M_{P}}\right) ^{3}\left( \frac{T}{T_{c}-T_{N}}\right)
^{9}T_{c}^{3}.  \label{nb2}
\end{equation}

The bubbles thus nucleate during the short time $\delta t_{\Gamma
}$ about $t=t_{m}$, and expand with a velocity $v_1\simeq
v_{w}\left( t_{N}\right) $ for a time $ \delta t_{1}$, until the
temperature gets close to $T_c$. According to Eqs.
(\ref{vwlinear}) and (\ref{ti}),
\begin{equation}
v_1\simeq \frac{1}{\eta} \left(\frac{16\pi
\sigma^3}{3T_c}\right)^{1/2} K^{-1/2}, \label{vwn}
\end{equation}
where $K$ is a shorthand for the sum of $\log $s in (\ref{ti}). Notice that in Eq.
(\ref{vwlinear}), $v_w\propto L\left(T_c-T \right)$, but also $T_c-T_N \propto
L^{-1}$, so $v_w$ only depends on $L$ through the logs in $K$.

If a non-negligible fraction of the volume is taken up by bubbles during the first
stage, the interval $\delta t_{1}$ can be estimated from $\left( 4\pi /3\right)
v_{w}^{3}\left( t_{N}\right) \delta t_{1}^{3}n_{b}\sim 1$. This gives
\begin{equation}
\delta t_{1}\sim \left( \frac{L^{4}}{\sigma ^{3}T_{c}^{7}}\right) ^{1/6}\left(
\frac{S_{3}\left( T_{m}\right) }{T_{m}}\right) ^{-1/2}\left( t_{N}-t_{i}\right) .
\label{deltat1}
\end{equation}
So, apart from a model-dependent factor, we find a general
tendency to the relations $\delta t_\Gamma\ll \delta t_{1}<
t_{N}-t_{i}$. The value of $S_{3}\left( T_{m}\right) /T_{m}$ (and
that of $T_{m}$) can be obtained similarly to the case of $T_{N}$.
It is interesting to note that $\delta t_{1}$ has only a
logarithmic dependence on the friction coefficient $\eta $. This
is because the dependence on the wall velocity is twofold. On one
hand, the lower the wall velocity, the longer the time $\delta
t_{1}$ needed to reheat the plasma. But on the other hand, the
lower the wall velocity, the longer will also be the time $\delta
t_{\Gamma }$ in which bubbles are formed, and the larger their
number. This causes a shorter $\delta t_{1}$, since there are more
bubbles to produce the reheating.

\subsection{The second stage: phase equilibrium}

\subsubsection{Inhomogeneous nucleation}

If the formation of bubbles is associated with the presence of impurities, the phase
transition occurs at $T\simeq T_{c}$, and the number density of bubbles $ n_{b}$ is an
external parameter that depends on the density of impurities. According to Eqs.
(\ref{scalepos})-(\ref{fbeqpos}), for small $r$ the evolution of the phase transition
is given by
\begin{equation}
f_{b}\simeq 3Hr^{-1}\left( t-t_{i}\right) ,
\end{equation}
and the rate at which the phase transition goes on is $\dot{f}_{b}=3H/r$,
i.e., a factor of $1/r$ larger than the rate of expansion $3H$ of a comoving
volume. Our assumption $r\ll 1$ implies that $\dot{f}_{b}\gg H$ and $\delta
t\ll t$.

To calculate the velocity of the interfaces, we assume that all the bubbles begin to
expand at $t=t_{i}$. Thus, the fraction of volume occupied by bubbles is
$f_{b}=n_{b}\frac{4\pi }{3}R_{b}^{3}$, where $R_{b}\left( t\right) $ is the bubble
radius. At the midpoint of the transition we have $\dot{f}_{b}\sim 4\pi
n_{b}\bar{R}^{2}v_{w}$, where $\bar{R}=\left( 4\pi n_{b}/3\right) ^{-1/3}$ is the
average radius. Therefore the mean velocity is given by $v_{w}=\left( 4\pi
n_{b}/3\right) ^{-1/3}H/r$. We notice, however, that even in this case in which
$\dot{f}_{b}$ is constant, the wall velocity may change significantly during the
transition, since $v_w\propto R_b^{-2}$. The total variation of $v_w$ depends on
$n_b$.

\subsubsection{Homogeneous nucleation}

In the case of a phase transition with supercooling, the situation
is very similar after the plasma has reheated up to a temperature
$T_{r}\simeq T_{c}$ . The transition proceeds at a rate
\begin{equation}
\dot{f}_{b}\sim 4\pi v_{w}\left( t\right) \int_{t_{i}}^{t}\Gamma \left(
t^{\prime }\right) R^{2}\left( t^{\prime },t\right) dt^{\prime }.
\label{fdot2stage}
\end{equation}
In this stage the nucleation rate has turned off. We have seen
that $\Gamma $ peaks sharply at a certain time $t_{m}$ in the
previous stage, so we can write Eq. (\ref{fdot2stage}) as
$\dot{f}_{b}\sim 4\pi v_{2}n_{b}R^{2}\left( t_{m},t\right) $,
where $v_{2}\equiv v_{w}\left( T_{r}\right) $, $n_{b}$ is given by
Eq. (\ref{nb2}), and $R\left( t_{m},t\right) \simeq v_{2}\left(
t-t_{m}\right) $. In any case, the integral in (\ref{fdot2stage})
is an average of the squared radius of the bubbles, and for the
present estimations we can set $\dot{f}_{b}\sim 4\pi
v_{2}n_{b}\bar{R}^{2}$, with $\bar{R}$ given by $n_{b}$, just as
in the case of inhomogeneous nucleation. Since the temperature is
almost constant, $L\dot{f}_{b}\simeq 4\rho H$, i.e., all the
released latent heat is taken away by the expansion of the
Universe. Thus, again $\dot{f}_{b}=3H/r$ and the velocity
coincides with that of the inhomogeneous nucleation case
\begin{equation}
v_{2}\simeq\left( \frac{3}{4\pi n_{b}}\right) ^{1/3}\frac{H}{r}.
\label{vfinal}
\end{equation}
Although in Eq. (\ref{vfinal}) it seems that the wall velocity
during phase equilibrium does not depend on the friction, in fact
it is proportional to $\eta^{-1}$ due to the dependence of $n_b$.

Since $\dot{f}_{b}\sim H/r$, the duration of this stage is
\begin{equation}
\delta t_{2}\sim rH^{-1}.  \label{deltat2}
\end{equation}
The reheating temperature $T_{r}$ must be such that the pressure difference is
adjusted so as to give the velocity (\ref{vfinal}). Using Eq.~(\ref{vwlinear}), we
find
\begin{equation}
\frac{T_{c}-T_{r}}{T_{c}}=\frac{\eta H}{rL}\left( \frac{4\pi n_{b}}{3}\right) ^{-1/3}.
\end{equation}
As expected, the larger the latent heat, the closer will be
$T_{r}$ to $T_{c}$. Unlike $\delta t_2$, the values of $T_r$ and
$v_2$ are not easy to determine, since they depend on the number
of bubbles that nucleated in the previous stage. Using Eq.
(\ref{nb2}) and taking into account that $HM_P/T^2\sim 1$, we find
that roughly
\begin{equation}
T_c-T_r\sim \left( \frac{T_c-T_N}{T_c}\right)^2
\left(T_c-T_N\right),
\end{equation}
which confirms that generally, $T_c-T_r\ll T_c-T_N$. According to
Eq. (\ref{vwlinear}), the same relation holds for the velocities
$v_1$ and $v_2$.

\subsection{Coalescence}

In the range $0.3\lesssim f_{b}\lesssim 0.5$ bubble percolation takes place.
We have seen that this process gives a contribution to the bubble expansion
rate, of order $\left( \sigma /\rho R^{3}\right) ^{1/2}$. For $f_{b}\sim 1$
this rate is
\begin{equation}
\dot{f}_{\mathrm{coalescence}}\sim \left( \sigma /\rho n_{b}\right) ^{1/2}.
\label{ratecoal}
\end{equation}
To establish the importance of this rate we should compare it with the rates
$\delta t_{1}^{-1}$ or $\delta t_{2}^{-1}$, corresponding to the two stages
we have studied. Such a comparison is difficult to carry out without
specifying a model, so we will ignore this effect in the subsequent
discussions.

Although coalescence is bounded to occur in the above range of $f_b$, it could have
important consequences if the associated bubble growth rate is significantly larger
than those given by the time scales $\delta t_{1}$ and $\delta t_{2}$. In a specific
model, comparison of Eqs. (\ref{deltat1}) and (\ref{deltat2}) with (\ref{ratecoal})
should not be hard to do, once $n_{b}$ has been evaluated.

\section{Thermodynamical parameters \label{physquant}}

We have seen that all the parameters that describe the dynamics of
the phase transition (i.e. $\delta t_{1}$, $\delta t_{2}$, etc.)
depend on a few thermodynamical parameters, such as the latent
heat or the friction coefficient. The formation of the different
cosmological products of a phase transition thus depends on these
quantities, and also on other parameters, such as the conductivity
of the plasma. These quantities are physically related, since all
of them come from the equilibrium or non-equilibrium
thermodynamics of the same underlying theory. This should be taken
into account in phase transition calculations, when ranges of
parameters are considered. Unfortunately, it is hard in practice
to establish general relations between these physical parameters.
Depending on the theory, it may even be impossible to compute some
of these quantities.

One can gain some insight on the relations between thermodynamical
quantities by conveniently modelling the free energy. The problem
is further simplified by referring to the general form of the
perturbative effective potential. In that case the thermodynamical
quantities can be related to the parameters of the microscopic
theory. We dedicate this section and the following to study the
aforementioned relations. We will concentrate only in those
parameters which influence directly the dynamics of the phase
transition. An analysis of other parameters that affect the
generation of cosmological remnants is considered in \cite{cosmo}.

\subsection{Free energy and viscosity}

We assume the free energy density takes the form
\begin{equation}
\mathcal{F}=-\frac{\pi ^{2}}{90}g_{\ast }T^{4}+V\left( \phi ,T\right) +L,
\label{free}
\end{equation}
where the scalar field $\phi $ is the order parameter, and
\begin{equation}
V\left( \phi ,T\right) =D\left( T^{2}-T_{0}^{2}\right) \phi ^{2}-ET\phi ^{3}+
\frac{\lambda }{4}\phi ^{4}  \label{veff}
\end{equation}
is the free energy density difference between the symmetric and the broken-symmetry
phases. The parameter $g_{\ast }$ is the number of light species of the plasma. In
general, $g_{\ast }$ depends on temperature, but it is usually approximated by
\begin{equation}
g_{\ast }=\sum_{\textrm{bosons}}g_{i}+\frac{7}{8}\sum_{\textrm{fermions}}g_{i},
\label{g*}
\end{equation}
where the sums are on particles with masses $m_{i}<T$, and $g_{i}$
is the number of degrees of freedom of species $i$ (see appendix
\ref{ap1}).

It depends on each particular case wether Eqs.~(\ref{free}) and (\ref{veff}) can be
derived from the microscopic theory. At any rate, they can be regarded just as a
simple model for studying the dynamics of the phase transition, being the latter
first-order if the coefficient $E$ is nonvanishing. The parameters of $V\left( \phi
,T\right) $ can be chosen in such a way that the free energy carries the
thermodynamical properties of the theory we wish to study \cite{h95,hkllm93,m01}. For
instance, these parameters determine the values of the critical temperature, latent
heat, surface tension, and correlation length. The thermodynamical parameters could be
obtained, e.g., with lattice simulations (see for example \cite{lr01}). Then one can
use those values to calculate the parameters $T_{0}$, $ D$, $E$ and $\lambda $.
Furthermore, in general the order parameter $\phi $ is a Higgs field or a combination
of Higgs fields, and $V\left( \phi ,T\right) $ is the finite-temperature effective
potential (see e.g. \cite{q99}). We will consider this case in the next section.

The effect of viscosity on the propagation of the bubble wall is calculated by
considering its equation of motion in the hot plasma,
\begin{equation}
\square \phi +V^{\prime }\left( \phi \right)
+\sum_{i}g_{i}\frac{dm_{i}^{2}}{ d\phi }\int \frac{d^{3}p}{\left(
2\pi \right) ^{3}2E_{i}}f_{i}\left( k,x\right) =0,
\end{equation}
which can be derived by energy conservation considerations
\cite{lmt92,k92,dlhll92,mp95}. Here $V\left( \phi \right) $ is the zero temperature
effective potential, the sum is over all particles that couple to $\phi $, $m_{i}$ are
the $\phi $-dependent masses (see the appendices), and $f_{i}$ are the phase space
population densities. This equation can be obtained by thermally averaging the
operator equation for $\phi $. If we separate $f$ into the equilibrium population
$f_{0}$ plus a small deviation $\delta f$, we obtain the equation
\begin{equation}
\square \phi +V^{\prime }\left( \phi ,T\right) +\sum
g_{i}\frac{dm_{i}^{2}}{ d\phi }\int \frac{d^{3}p}{\left( 2\pi
\right) ^{3}2E_{i}}\delta f_{i}=0, \label{eqfiplasma}
\end{equation}
where $V\left( \phi ,T\right) $ is the finite temperature
effective potential, given by Eq.~(\ref{veff}). Since the
departure from equilibrium is proportional to the velocity of the
bubble wall, it is the last term in (\ref{eqfiplasma}) which gives
the friction force of the plasma.

A simple approach to the calculation of the wall velocity \cite{h95,hkllm93} consists
in replacing the last term in Eq. (\ref{eqfiplasma}) with a typical damping term of
the form $d\phi /dt$. Due to Lorentz invariance this term must be in fact of the form
$u^{\mu }\partial _{\mu }\phi $, where $u_{\mu }$ is the four-velocity of the plasma.
Eq.~(\ref{eqfiplasma}) then may be written as
\begin{equation}
\square \phi +V^{\prime }+\left( \tilde{\eta}T\right) u^{\mu }\partial _{\mu
}\phi =0,
\end{equation}
where $\tilde{\eta}$ is a dimensionless damping coefficient that depends on
the viscosity of the medium. Boosting to a frame that moves with the wall,
and assuming stationary and non-relativistic motion in the $z$ direction, we
have
\begin{equation}
\phi ^{\prime \prime }=V^{\prime }\left( \phi \right)
-\tilde{\eta} Tv_{w}\phi ^{\prime },
\end{equation}
where $\phi ^{\prime }\equiv d\phi /dz$. Multiplying both sides by $\phi
^{\prime }$ and integrating over $-\infty <x<\infty $ we obtain
\begin{equation}
\tilde{\eta}T\sigma v_{w}=V\left( T\right) ,
\end{equation}
where $V\left( T\right) $ is the free energy difference between the two phases,
defined in Eq. (\ref{deltav}), and $\sigma $ is the surface tension of the wall, given
by Eq. (\ref{formulasigma}). We have assumed here that temperature is constant across
the wall. This is right if the wall velocity is small enough, so that the latent heat
it releases has time to be uniformly distributed throughout space.

Hence, the pressure difference is equilibrated by a friction force proportional to the
wall velocity. The constant of proportionality is the friction coefficient $\eta
=\tilde{\eta}T\sigma$. Since the tension of the wall is related to the wall width
$L_w$ by $\sigma\simeq\phi_m^2/L_w$ [see the first integral in Eq.
(\ref{formulasigma})], we can also write $\eta =\tilde{\eta}T\phi_m^2/L_w$.

A shortage of modelling the viscosity of the plasma in this way is that $
\tilde{\eta}$ is a free parameter. The correct expression for $\eta $ can be derived
from Eq. (\ref{eqfiplasma}). In appendix \ref{ap2} we show that particles with a
thermal distribution give a friction coefficient
\begin{equation}
\eta_\textrm{th} \simeq \tilde{\eta}_\textrm{th}\frac{\phi
_{m}^{2}}{T}\sigma , \label{etath}
\end{equation}
while  the contribution of infrared gauge bosons is
\begin{equation}
\eta_\textrm{ir} \simeq \tilde{\eta}_\textrm{ir}\frac{T^3}{L_w} .
\label{etair}
\end{equation}
Evidently, both formulas agree with the above result if $\phi _{m}\sim T$. This
treatment also allows for the evaluation of the coefficients $\tilde{\eta}$, which
depend only on the particle content of the plasma.

\subsection{Thermodynamical quantities and phase transition dynamics}

The free energy given by Eqs. (\ref{free}) and (\ref{veff}) bears a first-order phase
transition, with two minima separated by a barrier. The critical temperature is
related to $T_{0}$ by
\begin{equation}
\frac{T_{c}^{2}-T_{0}^{2}}{T_{c}^{2}}=\frac{E^{2}}{\lambda D}.  \label{tct0}
\end{equation}
At $T> T_{c}$ the global minimum of the potential is  $\phi =0$. At the critical
temperature the two minima become degenerate, and below this temperature the stable
minimum is
\begin{equation}
\phi _{m}\left( T\right) =\frac{3ET}{2\lambda }\left[
1+\sqrt{1-\frac{8}{9} \frac{\lambda D}{E^{2}}\left(
1-\frac{T_{0}^{2}}{T^{2}}\right) }\right] \ . \label{fimin}
\end{equation}
At $T=T_{0}$ the barrier between minima disappears and $\phi =0$
becomes a maximum of the potential. Therefore the phase transition
occurs at some stage in between $T_{c}$ and $T_{0}$. The value
$\phi =0$ corresponds to the symmetric, high-temperature phase,
and $\phi \neq 0$ corresponds to the broken-symmetry,
low-temperature phase. The jump of the order parameter from the
high temperature phase to the low temperature one is thus
\begin{equation}
\phi _{m}\left( T_{c}\right) =2ET_{c}/\lambda .  \label{fimtc}
\end{equation}

According to Eqs. (\ref{free}) and (\ref{veff}), the free energy density of the
symmetric phase is
\begin{equation}
\mathcal{F}_{u}=-\frac{\pi ^{2}}{90}g_{\ast }T^{4}+L.
\end{equation}
This gives the equation of state of a hot relativistic plasma with
a positive cosmological constant
\begin{equation}
\rho _{u}=\frac{\pi ^{2}}{30}g_{\ast }T^{4}+L,\quad p_{u}=\rho _{u}/3-4L/3.
\end{equation}
The free energy of the broken-symmetry phase is $\mathcal{F}=\mathcal{F} _{u}+V\left(
T\right) $. The energy density of the broken-symmetry phase is $\rho _{b}=\rho
_{u}+\Delta \rho $, with
\begin{equation}
\Delta \rho =V\left( T\right) -TV^{\prime }\left( T\right) .
\end{equation}
The entropy density of the symmetric phase is $s_u=2 \pi ^{2}g_{\ast }T^{4}/45$, and
that of the broken-symmetry phase is $s_b=s_u -V'\left(T\right)$. The latent heat of
the phase transition is given by $L=\Delta \rho \left(T_{c}\right) =T_c\Delta
s\left(T_{c}\right)$; hence,
\begin{equation}
L=8D\left( \frac{E}{\lambda }\right) ^{2}T_{c}^{2}T_{0}^{2}. \label{l}
\end{equation}
Comparing with Eq. (\ref{fimtc}), we find the relation $ L= 2D\phi_m^2T_0^2$ between
the discontinuity of the order parameter and that of the energy density. As expected,
strongly first-order phase transitions (i.e., with large $\phi_m$) have large latent
heat.

The surface tension of the bubble wall is given by Eq. (\ref{formulasigma}) in the
thin wall approximation. At the critical temperature the effective potential is given
by
\begin{equation}
V\left( \phi ,T_{c}\right) =\frac{4\left( ET_{c}\right)
^{4}}{\lambda ^{3}} x^{2}\left( 1-x\right) ^{2},  \label{vtc}
\end{equation}
where $x\equiv \lambda \phi /2ET_{c}=\phi /\phi _{m}$. Hence, Eq. (\ref{formulasigma})
is easily  integrated and
\begin{equation}
\sigma \left(T_{c}\right) =\frac{2\sqrt{2}E^{3}}{3\lambda ^{5/2}}T_{c}^{3}.
\label{sigma}
\end{equation}
Although we haven't used it explicitly, in this approximation the field configuration
near the wall can be solved analytically with the help of Eq. (\ref{vtc}), and gives
the kink profile,
\begin{equation}
\phi \left( z\right) =\frac{\phi _{m}}{2}\left( 1+\tanh
\frac{z}{L_{w}} \right) , \label{kink}
\end{equation}
where
\begin{equation}
L_{w}=\phi _{m}^{2}/3\sigma  \label{width}
\end{equation}
is the wall width \footnote{$L_{w}$ may change during the bubble expansion due to the
friction with the plasma \cite{mp95}. We shall neglect this effect.}.

Using Eqs. (\ref{fimtc}), (\ref{sigma}) and (\ref{width}) we find the values of the
friction coefficients (\ref{etath}) and (\ref{etair}),
\begin{eqnarray}
 \nonumber
  \eta_\textrm{th} &=& \frac{8\sqrt{2}}{3}
\frac{E^{5}\tilde{\eta}_\textrm{th}}{\lambda ^{9/2}}T_{c}^{4}, \\
  \eta_\textrm{ir} &=&
\frac{E\tilde{\eta}_\textrm{ir}}{\sqrt{2\lambda}}T_{c}^{4}.
\label{etas}
\end{eqnarray}
The two contributions have different parametrical dependence, so each will dominate in
different regions of parameter space. For instance, if $E\ll \lambda$ the infrared
boson contribution may be much larger than that of thermal particles. The maximum
velocity of bubble walls occurs at $T\simeq T_N$. According to Eqs. (\ref{vwn}) and
(\ref{etas}), this velocity is the smallest among
\begin{equation}
    v_{\textrm{th}}\sim \frac{\lambda^{3/4}} {E^{1/2}
    \tilde{\eta}_\textrm{th}}K^{-1/2},\;
    v_{\textrm{ir}}\sim \frac{E^{7/2}} { \lambda^{13/4}
    \tilde{\eta}_\textrm{ir}}K^{-1/2}.
\end{equation}
To determine which one is the correct, it is necessary to know the relations between
the coefficients $E$, $\lambda$, and $\tilde{\eta}$.  We see that in the opposite
limiting cases $E\ll \lambda$ and $E\gg \lambda$, one of the two velocities is $\ll
1$, unless $\tilde{\eta}$ is too small. In the case $E\sim\lambda$, the wall velocity
will be small if one of the conditions $E^{1/4}\ll\tilde{\eta}_\textrm{th}$ or
$E^{1/4}\ll\tilde{\eta}_\textrm{ir}$ is fulfilled, which does not seem unlikely in
general (see the next section). This supports the assumption of non-relativistic wall
velocities.

It is evident that with the aid of the model (\ref{free},\ref{veff}) we can get more
information about the generalities of phase transition dynamics. For instance, if we
write Eq. (\ref{ti}) as a function of $E$, $D$, and $\lambda $, and compare with Eq.
(\ref{tct0}), then we can locate the nucleation temperature in the interval
$T_{c}-T_{0}$,
\begin{equation}
T_{c}-T_{N}\sim \frac{E^{1/2}}{\lambda ^{3/4}}K^{-1/2}\left( T_{c}-T_{0}\right) ,
\label{aproxtn}
\end{equation}
If $E$ and $ \lambda $ are comparable, this gives a value of $T_{c}-T_{N}$ roughly an
order of magnitude less than $T_{c}-T_{0}$.

The relations between the different quantities that determine the dynamics of the
phase transition are apparent in the above expressions. Specific relations will be of
interest for different cosmological consequences. As an example, let us consider the
effect of modifying the theory in order to obtain a more strongly first-order phase
transition. To do that, we have to enlarge the value of the order parameter. Assume we
accomplish this by increasing the value of the parameter $E$ and keeping the other
parameters invariant [see Eq.(\ref{fimtc})]. Then, there will be more supercooling,
and one expects a larger departure from thermal equilibrium, since the pressure
difference at $T=T_{N}$ will be larger. However, according to Eqs. (\ref{l}) and
(\ref{etas}), $L$ and $\eta $ also increase. This tends to decrease the wall velocity
in the two stages of the phase transition, in opposition to the effect of
supercooling.

In this work we assume for simplicity that $g_{\ast }$ remains constant throughout the
phase transition. In fact, the number of effectively massless degrees of freedom may
change during the phase transition. It is conceivable that some particles acquire
large masses and decouple from the thermal bath; then $\Delta g_{\ast }\equiv g_{\ast
u}-g_{\ast b}>0$. For instance, during the quark-hadron phase transition $g_{\ast }$
changes substantially. It is interesting to note that such a change may affect
considerably the dynamics of the phase transition, even in the case $\Delta g_{\ast
}\ll g_{\ast }$. The effect of a decrease of $g_{\ast }$ during the phase transition
is twofold. To begin with, the free energy of the broken-symmetry phase is larger than
in the case of constant $g_{\ast }$, so the critical temperature is lower (it is given
by $V\left( T_{c}\right) =-\pi ^{2}\Delta g_{\ast }T_{c}^{4}/90$ \cite{gw81,qcd}).
Therefore the phase transition is stronger \footnote{This could be important for
baryogenesis \cite{cosmo}.}, and the latent heat $T_{c}V^{\prime }\left( T_{c}\right)
$ is larger. In addition, the entropy released by the decoupling species gives an
extra contribution of $ 4\pi ^{2}\Delta g_{\ast }T_{c}^{4}/90$ to the latent heat.
This contribution is comparable to the value of $L$ as given by Eq.~(\ref{l}), if
$\Delta g_{\ast }\gtrsim D\left( E/\lambda \right) ^{2}\left( 1-E^{2}/\lambda D\right)
$. In the case of a perturbative effective potential, this condition may be easily
fulfilled for $\Delta g_{\ast }\sim 1$.

\section{The physical quantities in perturbation theory \label{perturb}}

If perturbation theory is applicable, the one-loop effective potential at
high temperature has oftentimes the form of Eq.~(\ref{veff}), with
parameters generally given by
\begin{eqnarray}
D &=&\sum_\textrm{bosons}\frac{g_{i}h_{i}^{2}}{24}+ \sum_\textrm{fermions}
\frac{g_{i}h_{i}^{2}}{48},
\notag \\
T_{0}^{2} &=&\frac{1}{D}\frac{m_{h}^{2}}{4},  \label{potparam} \\
E &=&\frac{2}{3}\sum_{\substack{\mathrm{gauge}  \\
\mathrm{bosons}}}
\frac{g_{i}h_{i}^{3}}{12\pi},  \notag \\
\lambda &=&m_{h}^{2}/2v^{2},  \notag
\end{eqnarray}
Here, $h_{i}$ are the couplings of the particles with $\phi $, $m_{h}$ is the Higgs
mass, and $v$ its zero temperature vev. The coefficient $E$ in general involves only
gauge bosons. In appendix \ref{ap1} we review the derivation of these results and
discuss on the general assumptions and approximations that lead to Eqs. (\ref{veff}),
(\ref{g*}), and (\ref{potparam}). In the discussions that follow we will sometimes
take the electroweak theory as a reference point. The parameter $T_{0}$ gives the
temperature scale of the phase transition. Its order of magnitude is determined by
$m_{h}$, so it may be quite less than the scale $v$ if $\lambda $ is small. Anyway,
for the dynamics of the phase transition, the difference $T_{c}-T_{0}$ is more
important than the temperature scale $ T_{0}$.

Regarding the viscosity of the plasma, we show in appendix
\ref{ap2} that the contribution of thermal particles to the
parameter $\tilde{\eta}$ is given by
\begin{equation}
\tilde{\eta}_\textrm{th}\simeq \sum 3\left( \frac{\log \chi
_{i}}{2\pi ^{2}}\right) ^{2}g_{i}h_{i}^{4}  \label{etatildeth}
\end{equation}
where $\chi _{i}=2$ for fermions and $\chi _{i}=h_{i}^{-1}$ for
bosons. Therefore the contributions of bosons to $\eta $ have an
enhancement of $ \left( \log h_{i}^{-1}/\log 2\right) ^{2}$ with
respect to fermions with the same Yukawa coupling. For instance,
for $h\sim 0.1$ the boson enhancement is $\sim 10$. This means
that friction may be much stronger in supersymmetric theories than
in non-supersymmetric ones. For instance, it was found in Ref.
\cite{js01} that a light stop may slow down the electroweak bubble
wall in the MSSM an order of magnitude with respect to the SM. The
enhancement is larger for lighter particles, but these do not
contribute to the friction due to the $h_{i}^{4}$ dependence.

The contribution of infrared gauge bosons is
\begin{equation}
\tilde{\eta}_\textrm{ir}\simeq \sum \frac{g_b \bar{g} h_b^2
}{32\pi}\log \left[ m_b \left( \phi_m\right)L_w \right].
 \label{etatildeir}
\end{equation}
Here, the sum is only on gauge bosons, but the coefficient $\bar{g}$ also involves a
sum over particle species (see appendix \ref{ap2}). Furthermore, the gauge coupling
appears only squared, which means less suppression. The $\log$ enhancement in this
case is $\simeq \log \left(h_b \phi_m^3/ \sigma \right) \sim \left(\log h_b
\lambda^{-1/2}\right)$.

It is important to compare the value of the parameter $E$ with the other parameters,
since $E$ is responsible for the first-order nature of the phase transition. We can
see in the formulas of the previous section that all the thermodynamical quantities
are proportional to some power of $E$, while the parameters $D$ and $\lambda $ usually
appear in the denominators. In the perturbative approach (\ref{potparam}), this
parameter is generally smaller than the others. This is because $E$ is a sum of gauge
couplings to the third power, weighted with gauge boson degrees of freedom, while $D$
involves squared couplings, and the sum is over all degrees of freedom. Regarding
$\lambda $, it can be comparable to $E$, but this constrains the value of the Higgs
mass.

The smallness of $E$ indicates a tendency of perturbative effective potentials to give
weakly first-order phase transitions \footnote{We are only discussing one-loop order
here. Things may be different at two loops \cite{e96}.}. This is apparent in the
dependence of the order parameter, $\phi _{m}/T\sim E/\lambda $, or in the temperature
interval in which the first-order phase transition can occur, $\left(
T_{c}-T_{0}\right) /T_{c}\sim E^{2}/\lambda D$. For example, in the case of the
electroweak phase transition, we have $E\sim 10^{-3}$ and $D\sim 10^{-1}$ for the
minimal Standard Model. If we take a non-realistic value for $\lambda \sim E$ to get
an order parameter of order $T$, we find a temperature range $ T_{c}-T_{0}\sim
10^{-2}T_{c}$. In the specific case of the electroweak phase transition, a small value
of the Higgs field $\phi _{m}\left( T\right) $ is undesirable for electroweak
baryogenesis. In general, if this parameter is too small, the perturbative
approximation breaks down (see appendix \ref{ap1}). In the electroweak case, the way
out is to consider extensions of the SM which provide additional bosons that
contribute to the parameter $E$ \cite{cw95,cqw96}.

In the previous section we found that if we increase the parameter
$E$ while keeping the others constant, then we get a stronger
phase transition and larger supercooling, but also larger values
of $\eta$ and $L$, which slow down the dynamics. It is evident
that if $E$ is augmented by adding particles to the theory,  the
value of $D$ enlarges too, giving an additional increase of the
latent heat $L$. If we add only a boson, the relative change will
be more appreciable in $E$, because there are only a few terms in
its expression, but if the boson comes together with several new
species (as in the case of supersymmetry), then the change of $D$
will be much more substantial. According to Eqs.
(\ref{etatildeth}) and (\ref{etatildeir}), the friction
coefficient will also increase significantly when adding bosons to
the theory.

If $E^{2}/\lambda D\ll 1$, then $T_{c}\simeq T_{0}$, and
$L/T_{c}^{4}\simeq 8D\left( E/\lambda \right) ^{2}$. It is
interesting to compare the value of $L$ with that of $\delta \rho
\equiv \rho \left( T_{c}\right) -\rho \left( T_{N}\right) $, to
assess the effect of reheating, as discussed in section
\ref{transicion}. Using Eqs. (\ref{aproxtn}) and (\ref{tct0}), we
may write
\begin{equation}
\delta \rho \simeq 4\frac{T_{c}-T_{N}}{T_{c}}\sim
K^{-1/2}\frac{E^{5/2}}{ D\lambda ^{7/4}}.
\end{equation}
Therefore,
\begin{equation}
\frac{L}{\delta \rho }\sim \left( \frac{30K^{1/2}}{\pi ^{2}g_{\ast }}\right)
\left(\frac{D^{2}}{E^{1/2}\lambda ^{1/4}}\right).
\end{equation}
The first factor is likely of order 1 and depends essentially on the energy scale of
the transition. The second factor is determined by the dynamics. It depends mainly on
$D$, and may vary considerably if we change the particle content of the theory.
Exemplifying again with the electroweak theory, $D$ can vary from $\sim 10^{-1}$ in
the SM to $D>1$ in the MSSM, so we pass from little reheating in the first case to
large reheating in the latter. We remark that things may be quite different if Eqs.
(\ref{potparam}) are not valid. For instance, in the case of the quark-hadron phase
transition $L$ and $\delta \rho $ are typically of the order of the energy density
$\rho $.

\section{Conclusions \label{conc}}

In this paper we have performed an entirely analytical study of
first-order phase transitions in the radiation-dominated era. We
have seen that typically the high-temperature phase is supercooled
to a temperature $T_N$, after which the transition proceeds in two
steps, as sketched in Fig. \ref{esquema}. The first stage is
complex, and some rough approximations must be made for an
analytical treatment. Nevertheless, it can be checked with
numerical results, (e.g. \cite{m01}), that the orders of magnitude
are correct. The second stage is much more simple, since bubble
nucleation has effectively stopped and bubbles expand very slowly.
This stage develops very close to the critical temperature, with
almost zero pressure difference between the two phases. Therefore,
this part of the evolution is similar to the case of inhomogeneous
nucleation, in which the presence of impurities induces bubble
nucleations without need of supercooling.

We have studied the case of a phase transition at phase
equilibrium in some detail, taking advantage of the fact that it
can be solved analytically for any value of the relevant parameter
$r=L/T_cs(T_c)\sim L/\rho (T_c)$. This approach has allowed us to
calculate the back-reaction on the expansion rate $H$, which is
important for large $r$. It is well known that supercooling may
lead to exponential expansion of the Universe \cite{gw81,gw83}.
This possibility has been considered not only in the context of
inflationary models, but also for the quark-hadron phase
transition \cite{bcv00}. Although our approximations break down
for $L\gtrsim \rho(T_c)$, we observe the manifestation of the
energy of the false vacuum for large $L$. Even if the phase
transition begins at $T=T_c$, when $L$ is comparable to the energy
density of the plasma the transition may take a long time to
complete due to vacuum energy dominance.

For the more probable case of a phase transition with variation of
temperature, we have given a derivation of the
integro-differential equations that govern the dynamics. In
particular, we have found a simple algebraic relation between the
temperature, the fraction of volume occupied by the
low-temperature phase, and the scale factor of the Universe, which
holds under the usual assumption of adiabatic expansion. Using the
thin wall approximation and the linearized form of the free energy
difference, we have found analytical formulas for all the
quantities that characterize the dynamics of the phase transition
and may be relevant for the determination of its cosmological
consequences. These parameters are the durations $\delta t_1$ and
$\delta t_2$ of the two stages of the transition, the wall
velocities $v_1$ and $v_2$ at each stage, the total number density
of bubbles, the time interval $\delta t_\Gamma$ in which bubble
nucleation is active, etc. We have expressed these quantities in
terms of those that determine the dynamics, namely,
thermodynamical parameters like the latent heat $L$, the wall
tension $\sigma$, or the friction coefficient $\eta$. As expected,
for the phase equilibrium stage we have simple expressions, of the
sort $\delta t_2 \simeq r H^{-1}$, with obvious interpretation.
More complex formulas arise instead for the reheating stage.
Although these parameters must be calculated in each particular
case, some relations can be established, that hold quite broadly.
They allowed us to confirm some natural premises for the dynamics,
e.g., that $\delta t_\Gamma\ll \delta t_1\ll\delta t_2$ and
$v_2\ll v_1$.

We have studied also the interrelations between the
thermodynamical parameters. When necessary, we made use of a
simple model for the free energy. Aside from reproducing the
desired features of the phase transition, it is well known that
this model corresponds to the simplest high-temperature effective
potential that arises in perturbation theory. We have derived
general expressions for the parameters of this potential, which is
useful in establishing further relations between the
thermodynamical parameters. We have also derived general
expressions for the friction on the bubble walls. This is caused
by the perturbation from equilibrium of the particles of the
plasma due to the motion of the interfaces. We have compared the
case of thermal particles to that of coherent infrared bosons. We
have found a different parametric dependence of each contribution,
which indicates that each of them will dominate in different
parameter ranges. We have argued that probably one of these
contributions will cause the wall to move non-relativistically.
This justifies the near-equilibrium approximations that simplify
the analysis of the phase transition.

We have thus been able to find some general relations between the
parameters that determine the dynamics. For instance, we have seen
that if the first-order phase transition is strengthened, then the
supercooling is intensified, but also the friction and latent heat
are generically enlarged, giving a slower evolution. This general
feature is easily detected with ad hoc variations of the
parameters of the free energy, and further confirmed by the
relations that arise for perturbation-theory values of these
parameters. The amount of supercooling is characterized by the
difference between the energy density of the plasma at the
critical temperature, and that at which nucleation begins,
$\delta\rho=\rho\left( T_c \right)- \rho\left( T_N \right)$. The
specific relation between $\delta\rho$ and the latent heat is
decisive for the phase transition dynamics. As we have seen, the
ratio $L/\delta\rho$ can either be small or large, depending on
the theory. Of course, it is larger for stronger phase
transitions. We have argued that the case of interest corresponds
to a latent heat comparable to $\delta\rho$. On one hand, a small
$L$ is related to too weakly first-order phase transitions. On the
other hand, for large $ L$ most of the phase transition occurs
close to phase equilibrium, and can be described as a phase
transition at constant $T\simeq T_c $. This includes the case
$r\sim 1$ if $\delta \rho\ll \rho$, thus justifying the
approximation $r\ll 1$ in the case of supercooling.

For the study of the different cosmological consequences,
additional specific relations will be relevant in each case
\cite{cosmo}, which can be obtained from the present analysis. Our
analytical approximations will thus prove useful to include
details of phase transition dynamics in the calculation of cosmic
remnants, particularly with regard to the variation of the
pertinent parameters throughout the phase transition. For example,
the importance of the phase equilibrium stage has been already
investigated in \cite{h95,m01} in the context of electroweak
baryogenesis. Other cosmological consequences are affected as
well. For instance, the fact that the nucleation rate is turned
off due to reheating evidently modifies the number of nucleated
bubbles, and thus the density of topological defects.

\section*{Acknowledgements}

I am very grateful to Jaume Garriga for useful discussions and
suggestions. I wish to express my gratitude to Professor Luis
Masperi, who stimulated my interest on this subject. This work was
partially supported by a postdoctoral fellowship of the Ministerio
de Educaci\'on y Cultura, Spain.

\appendix

\section{Perturbative free energy \label{ap1}}

Following \cite{q99}, we will obtain the high-temperature effective potential (or free
energy) in the one-loop approximation, including leading-order plasma effects.
Additional terms appear at higher-loop order. For instance, potentially important
terms of the form $\phi ^{2}\log \phi $ arise at two loops \cite{e96}. However,
inclusion of two-loop corrections makes the situation more complicated and lies out of
the scope of our general analysis.

We consider a gauge theory which is spontaneously broken by a vev of a
scalar field $\phi $. The tree-level potential is
\begin{equation}
V_{0}\left( \phi \right) =-\frac{m^{2}}{2}\phi ^{2}+\frac{\lambda }{4}\phi
^{4},
\end{equation}
so the vev is given by $v^{2}=m^{2}/\lambda $ and the Higgs mass
by $ m_{h}^{2}=2\lambda v^{2}$.

At one-loop the effective potential picks up zero-temperature and
finite-temperature corrections. With a cut-off regularization and tree level
values for $v$ and $m_{h}$, the zero-temperature contribution of a particle
species is \cite{ah92}
\begin{equation}
\pm \frac{g}{64\pi ^{2}}\left\{ m^{4}\left( \phi \right) \left[
\log \frac{ m^{2}\left( \phi \right) }{m^{2}\left( v\right)
}-\frac{3}{2}\right] +2m^{2}\left( v\right) m^{2}\left( \phi
\right) \right\} ,
\end{equation}
where the $\pm $ is for bosons (fermions), $g$ is the number of degrees of
freedom of the species, and $m\left( \phi \right) $ is the mass of the
particle in the presence of the background scalar field. It is in general of
the form
\begin{equation}
m^{2}\left( \phi \right) =\mu ^{2}+h^{2}\phi ^{2},  \label{mdefi}
\end{equation}
where $h$ is the coupling of the particle with $\phi $ (i.e., the Yukawa
coupling, gauge coupling, etc.). The finite-temperature corrections are of
the form
\begin{equation}
\pm \frac{g}{2\pi ^{2}\beta ^{4}}J_{B,F}\left( m^{2}\left( \phi \right)
\beta ^{2}\right) ,
\end{equation}
where the functions $J_{B}$ and $J_{F}$ can be expanded in powers of $m/T$ for $m\ll
T$, and fall off exponentially for large $m/T$. Therefore species with $m\gg T$
decouple from the plasma and we make a high-temperature approximation in which we
consider only particles with $m<T$. Expanding up to $\mathcal{O}\left( m/T\right)
^{4}$ we have (see e.g. \cite{q99})
\begin{eqnarray}
\mathcal{F}\left( \phi,T \right) &=&constant-\frac{\pi ^{2}}{90}g_{\ast
}T^{4}-\frac{m_{h}^{2}}{4}\phi ^{2}  \label{veffap} \\
&&+\sum_{b}g_{b}\left( \frac{m_{b}^{2}}{32\pi ^{2}}+\frac{T^{2}}{24}\right)
m_{b}^{2}\left( \phi \right) +\sum_{f}g_{f}\left( -\frac{m_{f}^{2}}{32\pi
^{2}}+\frac{T^{2}}{48}\right) m_{f}^{2}\left( \phi \right)  \notag \\
&&-\frac{T}{12\pi }\sum_{b}g_{b}m_{b}^{3}\left( \phi \right)  \notag \\
&&+\frac{\lambda }{4}\phi ^{4}-\sum_{b}\frac{g_{b}}{64\pi
^{2}}\log \left( \frac{m_{b}^{2}}{T^{2}A_{b}}\right)
m_{b}^{4}\left( \phi \right) +\sum_{f} \frac{g_{f}}{64\pi
^{2}}\log \left( \frac{m_{f}^{2}}{T^{2}A_{f}}\right)
m_{f}^{4}\left( \phi \right) ,  \notag
\end{eqnarray}
where $g_{\ast }$ is the effective number of relativistic degrees of freedom, given in
Eq. (\ref{g*}) and $m_{i}\equiv m_{i}\left( \phi =v\right) $ are the physical masses.
We assume that particles contributing to $ \mathcal{F} $ do not decouple during the
phase transition, i.e., that the condition $m\left( \phi \right) \ll T$ is preserved
in the range of temperatures of interest. This is a reasonable assumption provided
that $\phi \ll v$ for temperatures close to $T_{c}$, which is consistent with the
one-loop approximation. If some particles decouple from the plasma during the
transition, the main effect is a change in $g_{\ast }$.

The $m^{3}$ term is the contribution of the bosonic zero modes to the one-loop
effective potential. This term is the most important to us; without it the phase
transition would be of second order. However, for the zero modes the loop expansion
has an infrared problem. The perturbative expansion breaks down at higher-loop order,
since higher loops contribute powers of $\alpha =h^{2}T^{2}/m^{2}\left( \phi \right) $
and of $\beta =h^{2}T/m\left( \phi \right) $ \cite{q99}. The way out is to dress the
zero modes with daisy and superdaisy diagrams. The result of this resummation is a
contribution of the form (to all order in $\alpha $ and to $\mathcal{O} \left( \beta
\right) $)
\begin{equation}
-\frac{T}{12\pi }\sum_{b}g_{b}\mathcal{M}_{b}^{3}\left( \phi \right)
\end{equation}
plus contributions proportional to $\phi ^{2}$ which are unimportant within
the present approximations. Thus the mass $m_{b}$ gets replaced with its
Debye mass
\begin{equation}
\mathcal{M}_{b}^{2}=m_{b}^{2}\left( \phi \right) +\Pi _{b}\left( \phi
,T\right) ,  \label{debye}
\end{equation}
where $\Pi _{b}$ is the self-energy of the boson particle. In general it is
a combination of squared coupling constants times $T^{2}$. The exception are
the transverse components of the gauge bosons, for which $\Pi =0$.

If we replace the Debye mass (\ref{debye}) in the cubic term of
Eq. (\ref{veffap}), and the masses (\ref{mdefi}) everywhere else,
the resulting terms can be grouped as follows:

\paragraph{Constant terms.}

These are contributions to the cosmological constant. The total cosmological constant
must be set by hand, so that it is almost zero after the phase transition.

\paragraph{$T$-dependent, $\protect\phi $-independent terms.}

Apart from the first term in (\ref{veffap}), there are also $T^{2}$ and logarithmic
terms, but these are of order $\left( \mu _{i}/T\right) ^{2}$ and $\left( \mu
_{i}/T\right) ^{4}$ with respect to the $T^{4}$ term, so we can neglect them within
the approximation $m_{i}\ll T$. We notice, however, that we could have a large
negative $\mu _{i}$ of order $T$, such that $m_{i}$ is small (see below). In any case,
these corrections modify the equations of state of both phases in the same way, and we
do not expect them to affect significantly the dynamics of the phase transition.

\paragraph{$\protect\phi ^{2}$ terms.}

The coefficient of $\phi ^{2}$ is the sum of a term proportional to $T^{2}$, a
constant term $m_{h}^{2}/4$, and other constant and logarithmic terms, which are $\sim
h_{i}^{2}m_{i}^{2}/32\pi ^{2}$. The latter are suppressed unless the Higgs mass is too
small, so we will disregard them. In any case, these corrections are inconsequential
for our purposes. They contribute to the value of the characteristic temperature
$T_{0}$ of the phase transition. However, for the dynamics of the phase transition the
precise value of $T_{0}$ is not relevant; the important parameter is the relative
difference between this temperature and the critical one, $\left( T_{c}-T_{0}\right)
/T_{0}$, which is independent of $T_{0}$.

\paragraph{$\protect\phi ^{3}$ terms.}

The $\mathcal{M}^{3}$ term  has contributions from all the bosons, proportional to
\begin{equation}
\left( h_{b}^{2}\phi ^{2}+\mu _{b}^{2}+\Pi _{b}\right) ^{3/2}.
\label{cubicterm}
\end{equation}
These terms may strongly affect the nature of the phase transition,
depending on the value of $\mu _{b}^{2}+\Pi _{b}\left( T_{c}\right) $. There
are two limiting cases,

\subparagraph{i)} if $\protect\mu _{b}^{2}+\Pi _{b}\left( T_{c}\right) \simeq 0$,
(\ref{cubicterm}) contributes a term of the form $T\phi ^{3}$ to the free energy,
which favors a strongly first-order phase transition, and \subparagraph{ii)} if
$\protect\mu _{b}^{2}+\Pi _{b}\left( T_{c}\right) \gg h_{b}^{2}\protect\phi ^{2}$, we
can expand (\ref{cubicterm}) in powers of $\phi $. This gives higher-order corrections
to the coefficients of $\phi ^{2}$ and $\phi ^{4}$, and no contribution to $\phi
^{3}$.

\medskip

Almost all particles fall in the second case, since $\Pi $ is of order $ h^{2}T^{2}$,
and $\phi \ll T$ near the phase transition. The transverse components of the vector
bosons, on the contrary, are protected against this thermal screening. Another
exception may be a scalar with a negative $\mu \sim T$. Such a particle would fall in
the first case or in an intermediate case, and may play a role in determining the
character of the phase transition \cite{cqw96}. However, such a tuning may induce
unwanted minima in the scalar potential \cite{cqw96,cw95}. We are not going to
consider this possibility here. Accordingly, the cubic term in the effective potential
is
\begin{equation}
-\frac{T}{12\pi }\sum \frac{2}{3}g_{b}h_{b}^{3}\phi ^{3}
\end{equation}
where the sum is only over gauge bosons, and the factor $2/3$ is
due to the fact that only two degrees of freedom of the massive
vector contribute\cite{dlhll92}.

\paragraph{$\protect\phi ^{4}$ terms.}

The corrections to $\lambda $ depend logarithmically on $T$, so the effective value of
$\lambda $ may be regarded to be constant during the phase transition. Furthermore,
these corrections are of order $ h_{i}^{4}/64\pi ^{2}$, so they can be neglected
provided that $\lambda \gtrsim h_{i}^{2}$. For simplicity we will assume that this is
the case.

\medskip

Putting all these terms together we see that, under the above assumptions and
approximations, the free energy density takes the form displayed in Eqs.
(\ref{free},\ref{veff}), with coefficients given by Eqs. (\ref{potparam}).

\section{Friction coefficient \label{ap2}}

In this appendix we make a derivation of the friction exerted by the hot plasma on the
bubble walls. For that, we must calculate the departure from equilibrium of the phase
space population functions, $\delta f$ in Eq.~(\ref{eqfiplasma}). The friction on the
wall has been extensively studied in the case of the electroweak phase transition
\cite{lmt92,dlhll92,k92,mp95,js01,mt97,moore}. Our aim here is to discuss the general
dependence of the friction on the particle content of a theory, so we will need to use
some approximations in order to keep the description as general as possible.

\subsection{Fluid approximation}

We begin by considering the contribution of particles with $p\gg L_{w}^{-1}$ ($L_{w}=$
wall width), for which the background field varies slowly and the semiclassical (WKB)
approximation is valid. Since in general $L_{w}^{-1}\gg T$, this condition is
satisfied for all but the most infrared particles \cite{mp95}, which we study below.
We follow Refs.~\cite{mp95,js01}, but we use a simpler ansatz for the deviations from
equilibrium distributions. This will suffice for our purposes. We assume that the
population density of a particle species in the background of the domain wall (that
moves along the $z$-direction) is governed by the Boltzmann equation
\begin{equation}
\left[ \partial _{t}+\left( \partial _{p_{z}}E\right) \partial _{z}-\left(
\partial _{z}E\right) \partial _{p_{z}}\right] f=-C\left[ f\right] ,
\label{boltzmann}
\end{equation}
where $E=\sqrt{p^{2}+m\left( z,t\right) ^{2}}$ is the particle energy, $\partial
_{p_{z}}E=p_{z}/E$ is the particle velocity, $-\partial _{z}E=-\partial _{z}\left(
m^{2}\right) /2E$ is the force on the particle, and $C\left[ f\right] $ is the
collision integral.

We use the ansantz $f=f_{0}\left( E/T-\delta \right) $, where
\begin{equation}
f_{0}\left( x\right) =\frac{1}{e^{x}\pm 1},  \label{f0}
\end{equation}
so the deviation from $f_{0}\left( E/T\right) $ is $\delta f=-f_{0}^{\prime
}\left( E/T\right) \delta $. Thus we obtain an equation for $\delta $ by
linearizing the Boltzmann equation. Keeping only terms of order $\left(
m/T\right) ^{2}$ we have
\begin{equation}
\left( \frac{1}{2ET}\partial _{t}m^{2}-\partial _{t}\delta + \frac{p_{z}}{E}\partial
_{z}\delta \right)f_{0}^{\prime } +C\left[ f\right] =0. \label{boltzlinear}
\end{equation}
The mass of the particle is a function of $z-v_{w}t$, and so is
the perturbation $\delta $ if we assume a stationary state. Thus
we make the replacements $\partial _{t}m^{2}=-v_{w}\left(
m^{2}\right) ^{\prime }$ and $\partial _{t}\delta =-v_{w}\delta
^{\prime }$, where the prime means derivative with respect to $z$.
We further simplify Eq.~(\ref{boltzlinear}) by making the
integration $\int d^{3}p/\left( 2\pi \right) ^{3}$. We obtain
\begin{equation}
c_{2}v_{w}\delta ^{\prime }-\Gamma \delta =c_{1}v_{w}m^{2\prime }/2T^{2},
\label{fluideq}
\end{equation}
where $c_{1}$ and $c_{2}$ are defined by the integrals
\begin{equation}
c_{1}\equiv -\frac{1}{T^{2}}\int \frac{d^{3}p}{\left( 2\pi \right)
^{3}E} f_{0}^{\prime },\quad c_{2}\equiv -\frac{1}{T^{3}}\int
\frac{d^{3}p}{\left( 2\pi \right) ^{3}}f_{0}^{\prime },  \notag
\end{equation}
and we have written the collision integral in the form \cite{mp95}
\begin{equation}
\int \frac{d^{3}p}{\left( 2\pi \right) ^{3}}\frac{C\left[ f\right]
}{T^{2}} =T\Gamma \delta .  \label{colterm}
\end{equation}
To lowest order in $m/T$ we have $ c_{1f}=\log 2/2\pi ^{2}$ and $ c_{2f}=1/12 $ for
fermions, and $c_{1b}=\log \left( T/m\right) /2\pi ^{2}$ and $ c_{2b}=1/6 $ for
bosons.

For each particle species $i$ we have a fluid equation of the form
(\ref{fluideq}). These equations are coupled through the collision
term (\ref{colterm}), and $\Gamma $ is in principle a matrix with
indices running over all particle species\footnote{In Refs.
\cite{mp95,js01} a more complex approximation was made, where the
perturbation $\delta $ is split up into three different
perturbations, $ \delta =\mu /T+E\delta T/T^{2}+p_{z}v/T$. In that
case there are three fluid equations for each particle species,
with different rates $\Gamma _{\mu }$, $ \Gamma _{T}$ and $\Gamma
_{v}$, and there arise additional constants $c_{3}$ and $c_{4}$.
Our approximation corresponds to considering only the term $\mu
/T$.}. However, only particles with large Yukawa couplings are
relevant for the friction force, since they have stronger
interactions with the bubble wall. In accordance with
\cite{mp95,js01}, in this appendix we call \textquotedblleft
heavy" these particles with large $h_{i}$. Notice however that
heavy particles with large mass $\mu _{i}$ in the unbroken phase
may be thermally decoupled and not contribute to the friction at
all. The remaining \textquotedblleft light" particles can be
treated as a common background perturbation $\delta _{bg}$. The
fluid equation for the background is simpler and can be solved to
eliminate $\delta _{bg}$. Moreover,  the heavy particles
primarily collide with the light particles, so direct coupling
between heavy species can be neglected. The effect of the light
background, though, once eliminated from the equations, is to
introduce a weak coupling between heavy particles. As a
consequence, heavy particles are only weakly coupled through the
background, and the non-diagonal terms of $\Gamma $ are suppressed
with respect to the diagonal terms by a factor $1/g_{\ast
\mathrm{light}}$, where $g_{\ast \mathrm{light}}$ is the number of
light species of the background (it is proportional to the heat
capacity of the plasma). We will therefore neglect non-diagonal
terms of $\Gamma $ in our analysis. Calculating the rates $\Gamma
$ is well beyond the scope of this work. They are of the form
$\alpha ^{2}\log \left( 1/\alpha \right) T$, where $\alpha $ is a
gauge coupling \cite{mp95}. We will assume that in general,
$\Gamma \lesssim 10^{-1}T.$

The r.h.s of Eq.~(\ref{fluideq}) is the source term of the equation. It is localized
at the bubble wall, so we expect the same for $\delta $. Therefore we have $\delta
^{\prime }/\delta \sim 1/L_{w}$. Normally, $L_{w}\gtrsim 10T^{-1}$, so if the wall
velocity is small, the first term on the l.h.s of Eq.~(\ref{fluideq}) is much less
than the second one and  can be neglected. With this approximation the equation has a
simple solution,
\begin{equation}
\delta =-v_{w}\frac{c_{1}m^{2\prime }}{2T^{2}}.  \label{delta}
\end{equation}

If we now insert $\delta f_i=-f_{0}^{\prime }\left( E_i/T\right) \delta_i $ in
Eq.~(\ref{eqfiplasma}) we obtain
\begin{equation}
\left( v_{w}^{2}-1\right) \phi ^{\prime \prime }+V^{\prime }\left( \phi
,T\right) +\frac{T^{2}}{2}\sum g_{i}c_{1i}\frac{dm_{i}^{2}}{d\phi }\delta
_{i}.
\end{equation}
Replacing the value of $\delta _{i}$ given by Eq.~(\ref{delta}),  multiplying times
$\phi ^{\prime }$, then integrating with respect to $z$, we get
\begin{equation}
-V\left( \phi _{m},T\right) =v_{w}\sum g_{i}\int \frac{c_{1i}^{2}\left( m_{i}^{2\prime
}\right) ^{2}}{4\Gamma }dz.
\end{equation}
The l.h.s. is the pressure difference between the two phases. It is equilibrated by a
friction force of the form $\eta v_{w}$. The friction coefficient is thus
\begin{equation}
\eta =\sum_{i=``heavy"}\frac{g_{i}h_{i}^{4}}{\Gamma }\int c_{1i}^{2}\phi ^{2}\phi
^{\prime 2}dz.  \label{eta}
\end{equation}
The coefficient $c_{1}$ for bosons depends on $m_i$, but it is easy to see that its
variation  with $z$ can be neglected, and we can make the approximation
\begin{equation}
c_{1b}=\log h_{b}^{-1}/2\pi ^{2}.
\end{equation}

To evaluate the integral in (\ref{eta}) we use the thin wall
approximation (\ref{thinwallapp}),
\begin{equation}
\int \phi ^{2}\phi ^{\prime 2}dz=\int_{0}^{\phi _{m}}\phi ^{2}\sqrt{2V}d\phi
.  \notag
\end{equation}
It is clear that this integral goes like $\phi _{m}^{2}\sigma $.
For the model (\ref{veff}), the integral is easily calculated
using Eq. (\ref{vtc}). It gives $\left( 3/10\right) \phi
_{m}^{2}\sigma $, so the friction coefficient is given by
\begin{equation}
\eta =\sum 3\left( \frac{\log \chi _{i}}{2\pi ^{2}}\right)
^{2}\frac{ g_{i}h_{i}^{4}}{\left( \Gamma _{i}/10^{-1}T\right)
}\frac{\phi _{m}^{2}\sigma }{T},
\end{equation}
where $\chi _{i}=2$ for fermions and $\chi _{i}=h_{i}^{-1}$ for bosons. According to
the arguments above, we will make the assumption that the parenthesis in the
denominator of the last equation is roughly $\sim 1$. This gives Eqs. (\ref{etath})
and (\ref{etatildeth}).

\subsection{Infrared bosons}

It has been shown  \cite{moore} that coherent gauge fields can
have important contributions to the friction. Following Ref.
\cite{moore}, we will  estimate the contribution of a gauge boson
to $\eta $. Infrared boson excitations must be treated classically
\cite{mt97}; furthermore, the dynamics of the soft fields is
overdamped by hard particles \cite{asy97}. As a consequence, the
equation for the population function is given by \cite{moore}
\begin{equation}
\frac{\pi m_{D}^{2}}{8p}\frac{df}{dt}=-E^{2}f+\mathrm{noise},  \label{damp}
\end{equation}
which comes from a similar equation for the amplitude of the
field. Here, $ m_{D}$ is the Debye mass, given by
$m_{D}^{2}\sim\bar{g}h^{2}T^{2}$, where, according to our previous
notation, $h$ is the gauge coupling, and $\bar{g}$ is roughly
proportional to the number of particles that couple to the gauge
field. Averaging over the noise, we get the restoring term $
-E^{2}\delta f$ in the rhs of Eq. (\ref{damp}). Since
$f=f_{0}\left( E/T\right) +\delta f$, and $\delta f\ll f_{0}$ for
small $v_{w}$, we can write
\begin{equation}
\delta f=-\frac{\pi m_{D}^{2}}{16pTE^{3}}f_{0}^{\prime
}\frac{dm^{2}}{d\phi } \phi ^{\prime }v_{w}.
\end{equation}

Inserting in (\ref{eqfiplasma}), multiplying by $\phi ^{\prime }$, and
integrating as before, we obtain
\begin{equation}
V\left( T\right) =-\sum_{\mathrm{gauge}}g_{b}\frac{v_{w}\pi
m_{D}^{2}}{8} \int dz\left( m_{b}^{2\prime }\right) ^{2}\int
\frac{d^{3}p}{\left( 2\pi \right) ^{3}}\frac{f_{0}^{\prime
}}{4pTE^{4}}.
\end{equation}
Since the momentum integral is infrared dominated, we can
approximate $ f_{0}^{\prime }\left( x\right) \simeq -1/x^{2}$, so
the momentum integral yields $T/32\pi ^{2}m_{b}^{4}$. With
$m_{b}^{2}=h_{b}^{2}\phi ^{2}$, we have
\begin{equation}
V\left( T\right)
=v_{w}\sum_{\mathrm{gauge}}\frac{g_{b}m_{D}^{2}T}{64\pi } \int
dz\frac{\phi ^{\prime 2}}{\phi ^{2}}.
\end{equation}
The last integral can be calculated using again the thin wall
approximation, Eqs. (\ref{thinwallapp}) and (\ref{vtc}),
\begin{equation}
\int_{0}^{\phi _{m}}\frac{d\phi }{\phi ^{2}}\sqrt{2V}=\frac{2
}{L_w}\int_{0}^{1}dx\frac{1-x}{x}.  \label{integral}
\end{equation}
There is a logarithmic divergence that must be cut off where the
approximations used in this derivation break down \cite{moore}.
Perturbation theory breaks down when $m_{b}\left( \phi \right)
\sim h_{b}^{2}T$, i.e., at $\phi /\phi _{m}\sim h_{b}\lambda /E$.
The kinetic theory description that leads to Eq.
(\ref{eqfiplasma}) breaks down when $m_{b}\left( \phi \right) \sim
L_{w}^{-1}$, i.e., at $\phi /\phi _{m}\sim \sqrt{\lambda
/h_b^{2}}$. The latter occurs first, so the $\log $ is cut off at
$m_{b}\left( \phi \right) L_{w}\sim 1$. In Ref. \cite{moore} it is
argued that the contribution of very infrared degrees of freedom
is subdominant, since their wavelength cannot resolve the
thickness of the wall. Hence, the integral in (\ref{integral})
gives to leading $\log$, $\log\phi_m/\phi=\log \left( m_{b}\left(
\phi_m \right)L_{w}\right)$, and the friction coefficient is
\begin{equation}
\eta =\sum_{\mathrm{gauge}}\frac{g_{b}m_{D}^{2}T}{32\pi L_{w}}\log
\left( m_{b}\left( \phi_m \right)L_{w}\right) ,
\end{equation}
which gives Eqs. (\ref{etair}) and (\ref{etatildeir}).

\end{document}